\documentclass[twocolumn]{aastex701}

\usepackage{multirow,bigdelim,booktabs}
\usepackage{rotating}
\usepackage{amsmath}
\usepackage{mathtools}
\usepackage{array}

\usepackage{CJK}

\usepackage{todonotes}
\usepackage{soul}
\DeclareUnicodeCharacter{2202}{\partial}

\newcommand{\ww}{\cite{woosley_evolution_1995}}
\newcommand{\clof}{\cite{chieffi_explosive_2004}}
\newcommand{\cltt}{\cite{chieffi_pre-supernova_2013}}
\newcommand{\nkt}{\cite{nomoto_nucleosynthesis_2013}}
\newcommand{\sk}{\cite{sukhbold_core-collapse_2016}}
\newcommand{\lc}{\cite{limongi_presupernova_2018}}

\shorttitle{Empirical Constraints on Supernova Yields and Delay Times from Dwarf Spheroidal Galaxies}
\shortauthors{Heiger et al.}

\graphicspath{{./}{}}

\begin{document}
\begin{CJK*}{UTF8}{gbsn}

\title{It's a matter of time: 
Empirical Constraints on Supernova Yields and Delay Times from Dwarf Spheroidal Galaxies}


\correspondingauthor{Mair\'ead~E.~Heiger}
\author[0000-0002-2446-8332]{Mair\'ead~E.~Heiger}
\affiliation{David A. Dunlap Department of Astronomy \& Astrophysics, University of Toronto, 50 St George Street, Toronto ON M5S 3H4, CA}
\affiliation{Dunlap Institute for Astronomy \& Astrophysics, University of Toronto, 50 St George Street, Toronto, ON M5S 3H4, CA}
\affiliation{Data Sciences Institute, University of Toronto, 17th Floor, Ontario Power Building, 700 University Ave, Toronto, ON M5G 1Z5, CA}
\email[show]{mairead.heiger@mail.utoronto.ca}

\author[0000-0002-4863-8842]{Alexander~P.~Ji}
\affiliation{Department of Astronomy \& Astrophysics, University of Chicago, 5640 S Ellis Avenue, Chicago, IL 60637, USA}
\affiliation{Kavli Institute for Cosmological Physics, University of Chicago, Chicago, IL 60637, USA}
\affiliation{NSF-Simons AI Institute for the Sky (SkAI), 172 E. Chestnut St., Chicago, IL 60611, USA}
\email{alexji@uchicago.edu}

\author[0000-0003-2573-9832]{Joshua~S.~Speagle (沈佳士)}
\affiliation{Department of Statistical Sciences, University of Toronto, 9th Floor, Ontario Power Building, 700 University Ave, Toronto, ON M5G 1Z5, CA}
\affiliation{David A. Dunlap Department of Astronomy \& Astrophysics, University of Toronto, 50 St George Street, Toronto ON M5S 3H4, CA}
\affiliation{Dunlap Institute for Astronomy \& Astrophysics, University of Toronto, 50 St George Street, Toronto, ON M5S 3H4, CA}
\affiliation{Data Sciences Institute, University of Toronto, 17th Floor, Ontario Power Building, 700 University Ave, Toronto, ON M5G 1Z5, CA}
\email{j.speagle@utoronto.ca}

\author[0000-0002-9110-6163]{Ting~S.~Li}
\affiliation{David A. Dunlap Department of Astronomy \& Astrophysics, University of Toronto, 50 St George Street, Toronto ON M5S 3H4, CA}
\affiliation{Dunlap Institute for Astronomy \& Astrophysics, University of Toronto, 50 St George Street, Toronto, ON M5S 3H4, CA}
\affiliation{Data Sciences Institute, University of Toronto, 17th Floor, Ontario Power Building, 700 University Ave, Toronto, ON M5G 1Z5, CA}
\email{ting.li@astro.utoronto.ca}

\author[0000-0002-1445-4877]{Alessandro Savino}
\affiliation{Department of Astronomy, University of California, Berkeley, Berkeley, CA 94720, USA}
\email{asavino@berkeley.edu}

\author[0000-0002-7393-3595]{Nathan~R.~Sandford}
\affiliation{David A. Dunlap Department of Astronomy \& Astrophysics, University of Toronto, 50 St George Street, Toronto ON M5S 3H4, CA}
\email{nathan.sandford@astro.utoronto.ca}

\author[0000-0001-6196-5162]{Evan~N.~Kirby}
\affiliation{Department of Physics and Astronomy, University of Notre Dame, Notre Dame, IN 46556, USA}
\email{ekirby@nd.edu}

\author[0000-0002-4739-046X]{Mithi~A.~C.~de los Reyes}
\affiliation{Department of Physics \& Astronomy, Amherst College, 25 East Drive, Amherst, MA 01002, USA}
\email{mdelosreyes@amherst.edu}

\author[0000-0002-4733-4994]{Joshua~D.~Simon}
\affiliation{Observatories of the Carnegie Institution for Science, 813 Santa Barbara St., Pasadena, CA 91101, USA}
\email{jsimon@carnegiescience.edu}


\begin{abstract}

The chemical abundances of a stellar population encode information about nucleosynthesis and its astrophysical sites, but this information is confounded by the specific star formation history of the host galaxy.
As a result, placing empirical constraints on supernova yields and timing using abundances has been very challenging.
We introduce a galactic chemical evolution model DLEIY that uses an observed star formation history and metallicity distribution to reduce these confounding factors.
Using a joint statistical model of the dwarf spheroidal galaxies Sculptor and Fornax, simultaneous constraints on population-averaged yields and galactic outflows are achieved with DLEIY, without fixing the absolute scale of nucleosynthetic yields.
The Fe yield from core collapse supernovae is consistent with existing theoretical yield models, while the measured Mg yield is a factor of 2-4 higher, corroborating previous suggestions that yield models may under-predict [Mg/Fe].
We also find that the rate of Type Ia supernovae is enhanced by about a factor of 5 relative to field galaxies, and the delay-time distribution goes as $\sim t^{-2}$, a much steeper relationship than that measured from supernova surveys ($\sim t^{-1.1}$).
These findings may suggest a metallicity dependence of the Type Ia rate and delay-time distribution.
\end{abstract}

\section{Introduction} \label{sec:intro}

Present-day chemical abundances in stars reflect the enrichment history of their environment.
Studies of galactic chemical evolution (GCE) attempt to wind back the clock to explain how these observed chemical abundances arose.
The magnitudes of and correlations between different chemical abundances encode the timescales of processes like star formation and stellar evolution, the sites of different nucleosynthetic processes, a record of the binary population, the initial mass function, the galaxy's merger history, gas infall and outflow, and more \citep{kobayashi_chemo-dynamical_2023}. 
GCE is thus a rich source of information on any one such process or phenomenon \citep[e.g.,][]{tinsley_stellar_1979,edvardsson_chemical_1993,matteucci_effect_2009,romano_chemical_2013,vincenzo_chemical_2014,bisterzo_galactic_2017,alexander_inhomogeneous_2023}.

Metal production in supernovae is one of the most significant of these processes in a galaxy's chemical evolution.
Metal production can be decomposed into the yield (how much of an element is produced) and the delay-time distribution (when it is returned to the interstellar medium).
The delay-time distribution (DTD) of an event describes the rate of that event after a stellar population is formed.
The DTD of core-collapse supernovae (CCSNe) is dictated by stellar lifetimes.
The DTD of Type Ia supernovae (SNeIa) depends on the nature of the SNIa progenitor(s) \citep[see][for a review]{maoz_type-ia_2012}.
The specifics of the binary population, such as the binary fraction, the initial mass function, and the separation distribution, also affect the DTD\footnote{Hereafter, `DTD' refers to the SNeIa DTD unless otherwise specified}, regardless of the progenitors.

Surveys like the Lick Observatory Supernova Search \citep[LOSS;][]{li_lick_2000, li_nearby_2011}, the Nearby Supernova Factory \citep{aldering_overview_2002}, the Sloan Digital Sky Survey-II Supernova Survey \citep{frieman_sloan_2008}, and All-Sky Automated Survey for Supernovae \citep[ASAS-SN;][]{shappee_man_2014,kochanek_all-sky_2017} have long established that there is a relationship between SNeIa and the properties of their host galaxies (and/or their local environment) \citep[e.g.,][]{mannucci_supernova_2005,li_nearby_2011,quimby_rates_2012,childress_host_2013,graur_loss_2017a, graur_loss_2017b, brown_relative_2019}.
However, it is less clear whether there is \emph{intrinsic} variation in the DTD and specific SNIa rate ($N_{\mathrm{Ia}}/M_{\star}$) with properties like mass, metallicity, or redshift, or whether observed trends can be attributed to other properties of the galaxies like the specific star formation rate \citep[e.g.,][]{sullivan_rates_2006, smith_sdss-ii_2012, kistler_impact_2013, brown_relative_2019, wiseman_rates_2021, ma_supernovae_2025}.
Because chemical evolution is sensitive to the form of the DTD, especially at early times, it can provide a complementary view to supernova surveys and a natural exploration of any environmental dependence.

In particular, local dwarf spheroidal galaxies (dSphs) are a powerful tool with which to approach this question and, more generally, to test theories or models of chemical evolution.
dSphs host ancient, lower metallicity stellar populations, with many forming a significant fraction of their stars at $z>5$ \citep{hurley-keller_star_1998,lee_star_2009,de_boer_star_2012, deboer_star_2012b,weisz_star_2014}.
They therefore provide a crucial link to the high-redshift universe, with the benefit of being able to study resolved stars.
These low-mass and gas-poor systems are more chemically primitive than more massive galaxies and have less complicated merger histories, so they are easier to model.
Their chemical abundances and enrichment histories have provided useful insight into hierarchical galaxy formation \citep[e.g.,][]{venn_stellar_2004,frebel_linking_2010,lee_reconstructing_2015,fernandez-alvar_chemical_2017}.

Metal production has been challenging to constrain using traditional GCE models, however, due to non-trivial degeneracies with the star formation history (SFH) of the galaxy and galactic outflows \citep{andrews_inflow_2017,johnson_dwarf_2023}.
While GCE models have been used to constrain the relative yields of different elements \citep[e.g.,][]{johnson_empirical_2023,weinberg_scale_2024}, the \emph{absolute} scale or normalization of yields remains much more uncertain.
In this work, we use a GCE model with two methodological departures from typical models that are intended to address these issues.
Firstly, the model, which we call DLEIY, uses the observed SFH and observed metallicity distribution function (MDF), rather than forward-modelling them.
Second, we jointly model two galaxies simultaneously.
These changes are intended to address these degeneracies, and they also allow us to test whether both galaxies are compatible with the same DTD and/or the same yields.

In Section \ref{sec:about}, we introduce the DLEIY model and compare it to traditional GCE models.
Section \ref{sec:likelihood} outlines the likelihood function and sampling procedure.
We also describe a statistical method of reconstructing the age-metallicity relationship, which is a cornerstone of DLEIY.
In Section \ref{sec:mocks}, we briefly demonstrate the performance of DLEIY on mock galaxies.
Section \ref{sec:apply} details the application of DLEIY to the dwarf galaxies Fornax and Sculptor.
Section \ref{sec:discussion} includes the results and discussion of the DTD and yield constraints and their implications.
We summarize our conclusions in Section \ref{sec:conclusion}.


\section{Model description}\label{sec:about}

\begin{figure*}
    \centering
    \includegraphics[width=\textwidth]{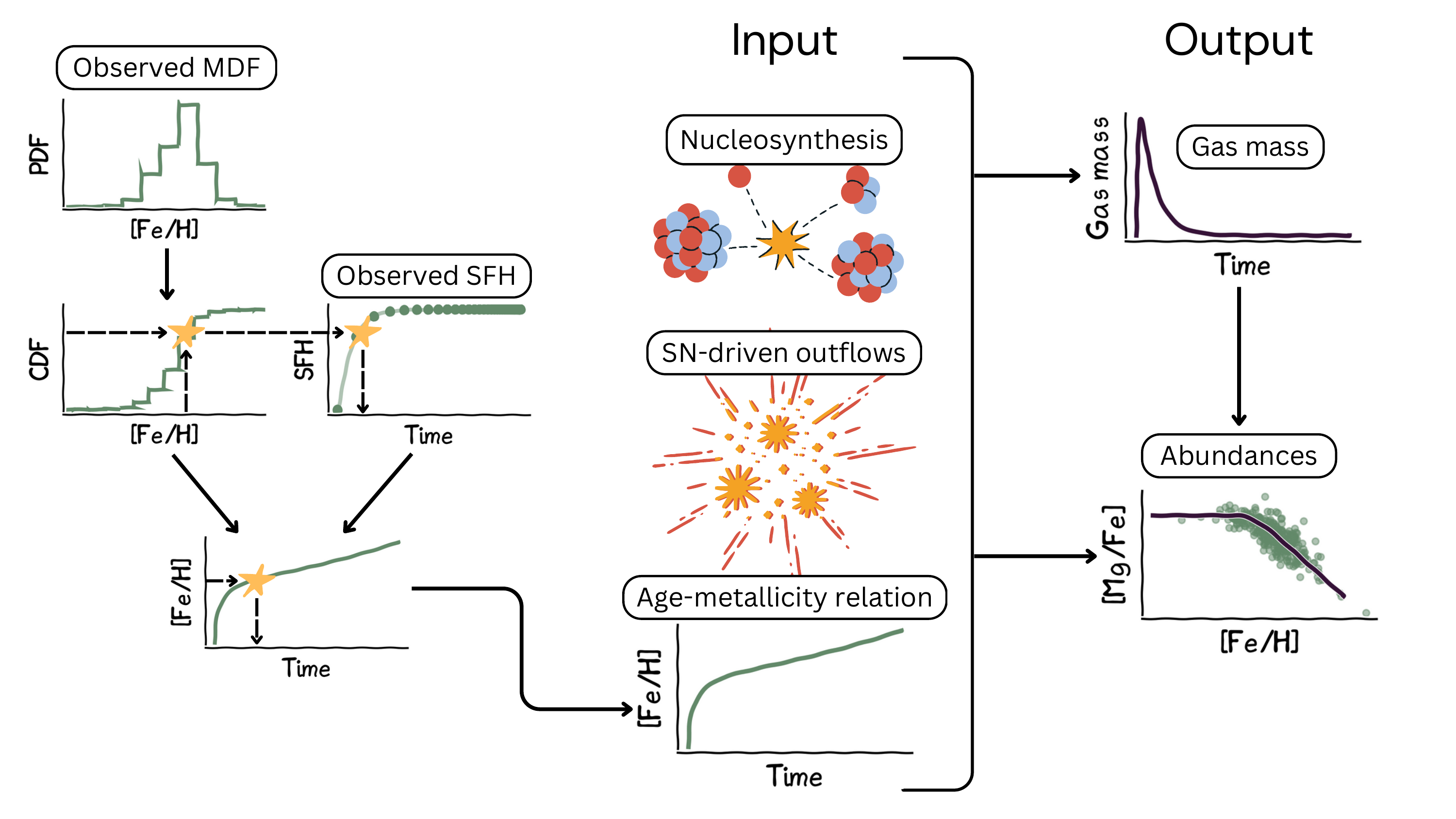}
    \caption{Schematic representation of DLEIY. (Left) The observed SFH and MDF are used to construct an age-metallicity relation. (Center) The age-metallicity relation, a parameterization of SN-driven outflows, and nucleosynthesis prescriptions (a delay-time distribution and population-averaged yields for different sources) are used to model the chemical abundances and gas mass (Right).
    \label{fig:schema}}
\end{figure*}

\begin{deluxetable*}{lccc}
\tabletypesize{\scriptsize}
\tablecaption{Summary of major differences between DLEIY and common approaches to GCE modelling, organized by the major physical processes/phenomena. This table is not intended to be absolute or exhaustive, but to contextualize DLEIY.
GCE models are very diverse, and some analyses include methods like sensitivity studies or model comparison that do not exactly map to the verbiage of `free' versus `fixed' parameters.
\label{tab:comparison}}
\tablehead{
\colhead{Process} &
\colhead{DLEIY} &
\colhead{Star formation-focused models}&
\colhead{Yield-focused models}
}
\startdata
Star formation rate & Fixed to observed & Parametric or implicit; usually free & Parametric; free or fixed \\
MDF            & Fixed to observed & Predicted & Predicted \\
Gas supply     & Implicit          & Parametric or implicit; usually free & Parametric or implicit; usually fixed \\
Yields         & Parametric; free  & Parametric; usually fixed & Choice between theoretical yield set or parametric; usually free \\
Delay-time     & Parametric; free  & Parametric; usually fixed & Parametric; usually fixed \\
Outflows       & Parametric; free  & Parametric; usually free & Parametric; usually fixed \\
\enddata
\end{deluxetable*}



DLEIY\footnote{DLEIY (pronounced ``delay") is named by reversing \emph{yield}, to emphasize the dual goal of constraining the delay-time distribution and nucleosynthetic yields} is a galactic chemical evolution model that replaces the parameters typically used to describe star formation with the observed MDF and observed SFH.
DLEIY is focused specifically on metal production---nucleosynthetic yields and the delay-time distribution---unlike many existing GCE models, which may aim to constrain or understand star formation or identify an optimal theoretical yield table.
This difference in scientific objective motivates a different methodology.
Figure \ref{fig:schema} shows a schematic diagram of DLEIY, and Table \ref{tab:comparison} summarizes some of the differences between DLEIY and other GCE models.

GCE models in the literature are diverse, but usually, the MDF and sometimes the SFH are forward modeled.
These models generally employ an analytic form of the star formation rate, gas mass, and/or the relationship between them \citep[e.g.,][]{schonrich_chemical_2009,romano_chemical_2013, vincenzo_chemical_2014,homma_new_2015, andrews_inflow_2017, weinberg_equilibrium_2017, cote_impact_2017, johnson_dwarf_2023}.
The creation and dilution of metals are directly related to the star formation rate and gas mass by some governing equation(s), so the observed abundances can be used to constrain them.


However, many dwarf galaxies in the Local Group have star formation histories that are poorly described by commonly employed analytic prescriptions, like the extended star formation of Fornax or the discrete bursts of Carina \citep{weisz_star_2014, de_boer_episodic_2014}.
Traditional GCE models often struggle to reproduce galaxies like these \citep[e.g.,][]{kirby_multi-element_2010,hasselquist_apogee_2021}.
Models that use an observed star formation rate are not uncommon \citep[e.g.,][]{lanfranchi_chemical_2003, de_boer_star_2012, homma_new_2015}, but these models typically still use a similarly restrictive parameterization/analytic form for the gas mass and star formation efficiency.
GCE models are also often under-determined, such that constraining both star formation and metal production is intractable. 
Typically, when models are used to constrain yields or evaluate theoretical yield tables, parameters related to star formation must be fixed \citep[e.g.,][]{romano_quantifying_2010,andrews_inflow_2017,palla_effects_2021,johnson_empirical_2023,liang_assessing_2023}, although this is not universally true \citep[e.g.,][]{philcox_optimal_2018,de_los_reyes_simultaneous_2022}.
Conversely, when models are applied to understanding star formation, yields and often the delay-time distribution are fixed \citep[e.g.,][]{kirby_multi-element_2011,weinberg_equilibrium_2017,andrews_inflow_2017,sandford_strong_2024,dubay_galactic_2024}.

To address these challenges, in DLEIY, the observed star formation rate (SFR) is used in lieu of a parametric form, and the gas mass is implicitly determined by an observed age-metallicity relation.
The only free parameters in DLEIY are those that describe yields, outflows, the delay-time distributions, and recycling of mass in star formation. 
The observed SFH and MDF replace typical GCE parameters like the infall timescale, star formation efficiency, star formation timescale, initial gas mass, etc.
DLEIY therefore does not make any predictions about star formation or gas infall; only metal production is forward-modeled.
Because the SFH is not restricted to a specific parametric form, DLEIY is very flexible. 
It is also more strongly informed by observational data, although its predictive scope is more limited as a result.

Despite these differences, the implementation of DLEIY is actually very similar to other GCE models, because the equations governing enrichment are the same, i.e., an accounting of sources and sinks of metals in the ISM.
Equation \ref{eq:sources} specifies the source terms in DLEIY: newly produced metals, and unprocessed metal mass returned to the ISM during stellar evolution (recycled mass).
Recycled mass is set by an instantaneous recycling rate $r$; we adopt a conventional value of $r = 0.4$ \citep{sandford_strong_2024}.
The recycled metal mass depends on the current abundance of that metal, $Z_{X}=M_{X}/M_{g}$.
The newly produced metal mass from an individual source equals the population-averaged yield ($y_{X}^{i}$) multiplied by the source rate.
The source rate is the convolution of the SFR and the DTD of the source $\Psi_i(t)$.
In this work, we consider only Fe from CCSNe and SNeIa, and Mg from CCSNe, although DLEIY is completely extensible to other elements and other sources, like asymptotic giant branch stars.

\begin{equation}
    \dot{M}_{X}^{in} = \overbracket{\sum_i y_{X}^{i} \times \underbrace{(\dot{M_{\star}}*\Psi_{i})(t)}_{\text{SN rate}}}^{\text{Metal mass produced by SNe}} + \overbracket{rZ_X \times \dot{M_{\star}}}^{\text{Recycled mass}}
    \label{eq:sources}
\end{equation}

Equation \ref{eq:sinks} specifies the sink terms in DLEIY: metals that are trapped in stars by recent star formation, and metals lost by stellar feedback, parameterized by a mass-loading factor $\eta$. 
The mass-loading factor is a constant proportion of the star-formation rate, where for every solar mass of gas formed into stars, $\eta$ of metals are ejected. 
This parameterization is typical in the literature for (semi)analytic GCE models.

\begin{equation}
    \dot{M}_{X}^{out} = \overbracket{Z_{X}\times \dot{M_{\star}}}^{\mathclap{\substack{\text{Metals in ISM} \\ \text{incorporated}\\ \text{into stars}}}}+ \overbracket{\eta Z_{X}\times \dot{M_{\star}}}^{\text{Outflows}}
    \label{eq:sinks}
\end{equation}

Equation \ref{eq:ode} combines Equations \ref{eq:sources} and \ref{eq:sinks} to give the change in metal mass in the ISM as a function of time.
This can be grouped into newly created metal mass, and metal mass (re-)processed by star formation.

\begin{equation}
    \dot{M}_{X} = \overbracket{\sum_i y_{X}^{i} \times (\dot{M_{\star}}*\Psi_{i})(t)}^{\text{Newly created metal mass}} - \overbracket{(1 + \eta-r)Z_X\times \dot{M_{\star}}}^{\mathclap{\substack{\text{Metal mass processed} \\ \text{by star formation}}}}\label{eq:ode}
\end{equation}

The gas supply in DLEIY is also different from typical prescriptions.
Some GCE models directly model gas inflows \citep[e.g.,][]{kirby_multi-element_2011,andrews_inflow_2017}; in others, it is parameterized using a star formation efficiency timescale \citep[e.g.,][]{cote_impact_2017}.
In DLEIY, the gas mass as a function of time is defined as the mass of hydrogen necessary to dilute the mass of Fe ($M_{\mathrm{Fe}}$) to the observed metallicity at a given time according to the age-metallicity relation (AZR):
\begin{equation}
    M_{g}(t) = \frac{M_{\mathrm{Fe}}}{Z_{\mathrm{Fe},\odot} }\times10^{-\mathrm{[Fe/H](t)}}
    \label{eq:azr}
\end{equation}

We assume that the stellar and gas-phase metallicity are the same.
The stellar metallicity is expected to be lower than the gas-phase metallicity \citep[e.g.,][]{gallazzi_ages_2005,lian_mass-metallicity_2018}, in which case the gas mass used here would overestimate the true gas mass.
We note that the magnitude of this offset in metallicity---and therefore the effect on the gas mass---is not clear, with estimates ranging from $<0.1$ dex \citep{zinchenko_different_2024} to $\sim0.8$ dex \citep{lian_mass-metallicity_2018}.

In practice, the SFH is first monotonically interpolated onto a fine time mesh using the shape-preserving Piecewise Cubic Hermite Interpolating Polynomial interpolant.
The observed SFR $\dot{M_{\star}}$ is given by the derivative of the SFH, multiplied by the total stellar mass formed (the present-day stellar mass corrected for high-mass stars that do not survive to present day).
The AZR is then constructed by mapping the empirical cumulative distribution function of the metallicity to the SFH by evaluating the percent point function of the MDF.
We assume instantaneous recycling for CCSNe ($\Psi_{\mathrm{CC}}=\delta(0)$) and a power-law DTD for SNeIa (see Table \ref{tab:params}), and calculate the SN rates by convolving the SFR and DTDs.
We then compute $M_{\mathrm{Fe}}$ and the gas mass using Equation~\ref{eq:azr}, as $\textrm{[Fe/H]}(t)$ is known per the observed AZR.
Then, we solve Equation \ref{eq:ode} for $M_{\mathrm{Mg}}$ by numerical integration (using the 5(4) Dormand-Prince method).
The abundance of Mg (or any other element) is given simply by $\mathrm{[Mg/H]}= \log_{10}{(M_{\mathrm{Mg}}/M_{g}})-\log_{10}{(Z_{\mathrm{Mg},\odot})}$, and $\mathrm{[Mg/Fe]=[Mg/H]-[Fe/H]}$.
The morphology and density of stars along the resulting ``track" through [Mg/Fe]-[Fe/H] space can be compared to observed abundances to constrain the yields, SNeIa DTD, and mass-loading factor.

We highlight two assumptions of this method: that the metallicity increases monotonically; and that the SFH and the MDF describe the same population.
The former is rarely assumed in GCE models, although it is often an outcome for single-zone models, and it is commonly assumed in SFH studies \citep[e.g.,][]{savino_hubble_2023,mcquinn_jwst_2024}.
This assumption could be broken if there is a late influx of pristine gas that dilutes the metal reservoir or a sudden change in outflows \citep{weinberg_equilibrium_2017}, but the star formation histories of Fornax and Sculptor do not suggest that this was the case.
Inhomogeneous mixing of the interstellar medium could also break one or both of these assumptions, as it introduces intrinsic scatter in the abundance distributions.
Using \citet{kirby_multi-element_2010} (the same dataset used in this work), \citet{escala_modelling_2018} found that the Si abundance distributions in Fornax and Sculptor are consistent with zero intrinsic scatter, so this effect is unlikely to be significant for this work.
With respect to the latter assumption, if the population sampled by the MDF is more metal-poor than that used for the SFH, then the age-metallicity relation will be systematically underestimated.
This could arise if there is a strong metallicity gradient, as the area covered by the SFH is small compared to the spectroscopic metallicity data.
See \cite{homma_new_2015} for further discussion of the effects of this type of area mismatch.
Relaxing these assumptions is an avenue for future development of the model.

\section{Likelihood and sampling procedure} \label{sec:likelihood}

\begin{figure*}
    \centering
    \includegraphics[width=\textwidth]{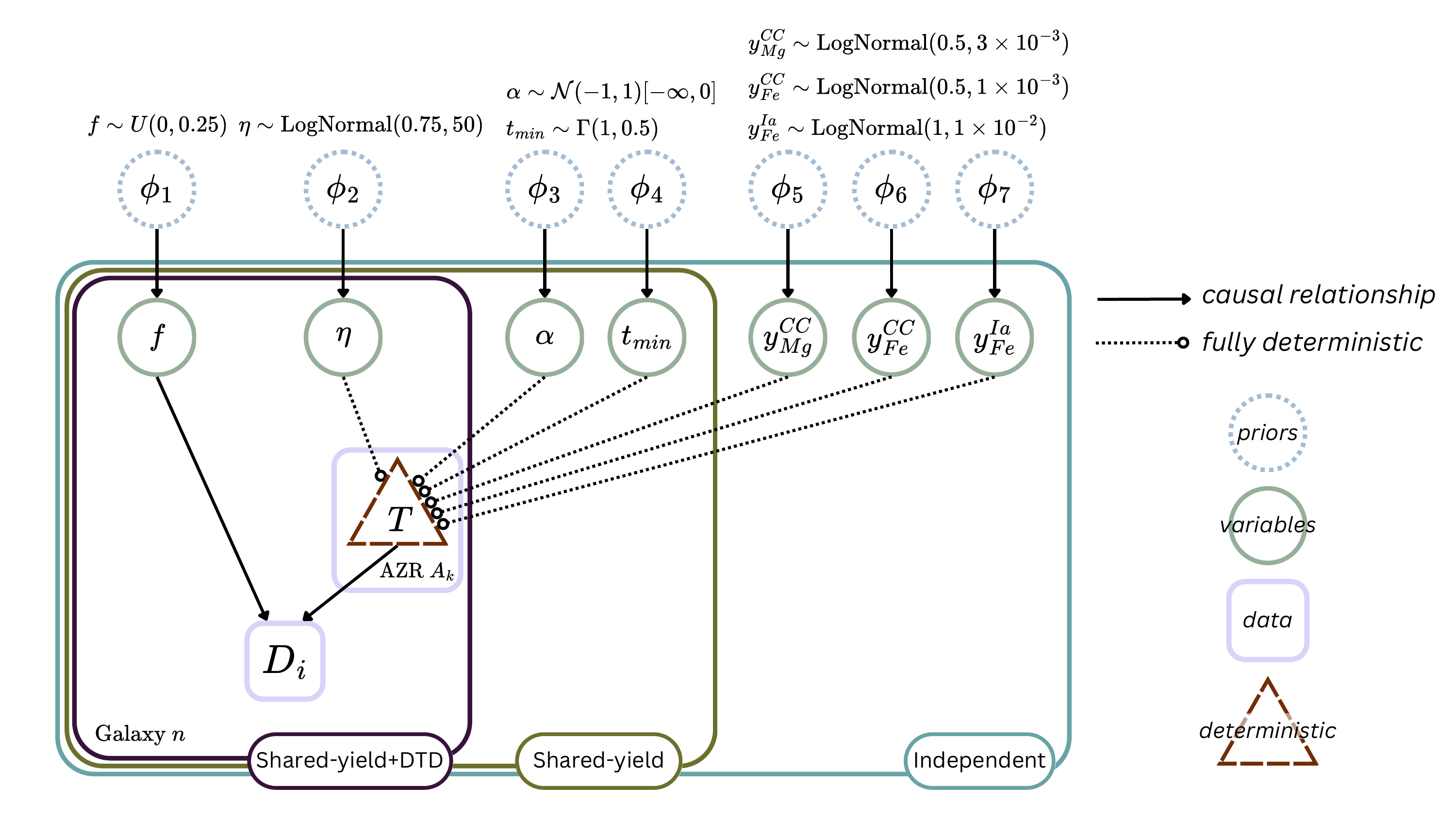}
    \caption{Graphical representation of the independent, shared-yield, and shared-yield+DTD models. The independent model has $7n$ parameters, where $n$ is the number of galaxies, as all parameters are unique to each galaxy. The shared-yield model has $4n+3$ parameters, and the shared-yield+DTD model has $2n+5$ parameters.}
    \label{fig:dag}
\end{figure*}

To fit DLEIY, we use a Bayesian approach.
The parameters that are fit and their priors are tabulated in Table \ref{tab:params}.
The priors are relatively broad and restrict the domain to physically meaningful values (i.e., $\eta$ must be positive).
We adapt the likelihood from \cite{johnson_dwarf_2023}, an inhomogeneous Poisson point process, and include an outlier model.
We also incorporate uncertainties in the SFH and MDF by averaging over an ensemble of AZRs, constructed from Monte Carlo draws of the SFH and MDF.
Uncertainty in the MDF is further addressed in Section \ref{sec:azr}; uncertainty in the SFH is addressed in Section \ref{sec:apply}.

Given parameters $\theta$ and an AZR draw $A_{k}$, DLEIY generates a model track $\{T_{j}\}=g(\theta,A_{k})$ through $\textrm{[Fe/H]-[Mg/Fe]}$ space.
Each datum $D_{i}$ is compared to each point $T_{j}$ along the model track, weighted according to how likely we are to observe a star at point $j$.
This weight is simply the SFR at the time point $j$ is reached (normalized to sum to unity).
In theory, this weight could also include a selection function.
For a draw of the AZR, $A_{k}$, the likelihood of observing a given datum $D_{i}=(\textrm{[Fe/H]}_{i}, \textrm{[Mg/Fe]}_{i})$ with covariance $C_{i}$ is:

\begin{multline}
    P(D_{i}\mid\theta,A_{k}) \propto \sum_{j=1}^{J} w_j~\mathcal{N}(D_{i}|T_{j},C_{i})= \\ 
    \sum_{j=1}^{J} w_j \exp\left( -\frac{1}{2} (D_{i}-T_{j}) C_i^{-1} (D_{i}-T_{j})^{\mathrm{T}}\right)
    \label{eq:inlier}
\end{multline}

\begin{equation}
w_j \propto \dot{M}_{\star,j}
\label{eq:weights}
\end{equation}

We combine this model with an outlier model.
Early investigations showed that inference was very sensitive to outliers, resulting in many local modes corresponding to models that fit one or a few points extremely well, at the expense of the others.
To reduce bias due to the influence of outliers, we adopt the outlier model described in \cite{hogg_data_2010}, which is a mixture model of an inlier and outlier data-generating process. 
There are many other approaches to outliers, such as clipping or non-Gaussian uncertainties, but this method is highly interpretable, as we can track the odds that a point is an outlier across the parameter space, and it does not exclude any particular solutions out of hand. 
We investigated removing outliers according to the local outlier factor \citep{breunig_lof_2000}, which also improved the multi-modality, but did not fully resolve it like the outlier model.
Clipping according to the local outlier factor yielded similar point estimates to the outlier model, with deviation being at the $1-2\sigma$ level.

The outliers are modeled as a uniform background in both $\mathrm{[Fe/H]}$ and $\mathrm{[Mg/Fe]}$.
The nature of this distribution is not especially important---the inclusion of \emph{any} outlier model will tend to improve the estimate, even one that is a poor match for the true outlier distribution.
We choose the bounds of the uniform distribution to be the bounds of the observed distribution +0.2 dex, so the probability of being an outlier $P_{\mathrm{out}}$ is constant.
The probability of being an inlier, $P_{\mathrm{in}}$, is given by Equation \ref{eq:inlier}. 
For a given draw of the AZR, $A_{k}$, and a mixture weight $f$, we can now write the likelihood as:

\begin{equation}
    \mathcal{L}_{k}\propto \prod_{i=1}^{N}(1-f)\times P_{\mathrm{in}}(\{D_{i}\}|~\theta,A_{k}) + f\times P_{\mathrm{out}}
\end{equation}

Each AZR is assumed to have equal weight, so the likelihood of observing a given datum $D_{i}$ for a given $\theta$ and ensemble of AZRs $\{A_{k}\}$ is a simple average over $k$:

\begin{multline}
    \log{\mathcal{L}(D_{i}|~\theta,\{A_{k}\})} \propto \\  
    \log{\sum_{k=1}^{K}(1-f)\times P_{\mathrm{in}}(D_{i}|\theta,A_{k})+f\times P_{\mathrm{out}}}\\-\log{K} 
\end{multline}

We implement three versions of this model, which are outlined in Figure \ref{fig:dag}. 
The first is the \emph{independent} model as described above: a model of a single galaxy, where every parameter is local to that galaxy.
The second is a \emph{shared-yield} model, a joint model of multiple galaxies where the galaxies have the same population-averaged yields, but other parameters (outflows and DTD) are local.
The third is a \emph{shared-yield, shared-DTD} model, where only outflows are local.
The nuisance parameter related to the outlier model, $f$, is local to each galaxy in all models.
In the shared-parameter models, the likelihood for each galaxy is independent, so the log-likelihoods are summed.


We sample from these models using \texttt{dynesty}, a Python package for nested sampling \citep{speagle_dynesty_2020,koposov_joshspeagledynesty_2024}.
Briefly, nested sampling estimates the Bayesian evidence $\mathcal{Z}$ by integrating the prior volume within shells of increasing likelihood (see \citealt{skilling_nested_2004,skilling_nested_2006} for more detail on nested sampling).
The evidence, which is difficult to obtain from sampling methods like Markov Chain Monte Carlo, is used for model comparison.
Nested sampling also provides posterior samples as a byproduct, which are used for parameter estimation like in other sampling algorithms.
We adopt a single ellipsoid as the bounding distribution \citep{mukherjee_nested_2006} and use random slice sampling \citep{neal_slice_2003,handley_polychord_2015a,handley_polychord_2015b}.
We use 500 live points and stop when $\Delta\ln{\mathcal{Z}_{i}} < 0.01$, meaning sampling stops when only $<1\%$ of the evidence remains.
We also tested uniform priors, as well as more permissive $\textrm{LogNormal}$ priors for the yields.
Agreement with different priors was around $1-2\sigma$, and the choice of prior does not affect any of the later conclusions.

\renewcommand{\arraystretch}{1.25}
\begin{deluxetable*}{llll}
\tablecaption{Parameters used in DLEIY and their prior distributions.\label{tab:params}}
\tablehead{
\colhead{Parameter} &
\colhead{Prior distribution} &
\colhead{Description} &
\colhead{Note}
}
\startdata
$y_{\mathrm{Mg}}^{CC}$ & $\mathrm{LogNormal}(0.5,3\times10^{-3})$ & Population-averaged CCSN Mg yield & $= \frac{m_{\mathrm{Mg}}}{\mathrm{CCSN}} \times \frac{\#\,\mathrm{CCSN}}{M_{\odot}\,\mathrm{formed}}$ \\
$y_{\mathrm{Fe}}^{CC}$ & $\mathrm{LogNormal}(0.5,1\times10^{-3})$ & Population-averaged CCSN Fe yield & --- \\
$y_{\mathrm{Fe}}^{Ia}$ & $\mathrm{LogNormal}(1,1\times10^{-2})$ & Population-averaged SNIa Fe yield & --- \\
\hline
$\alpha$ & $\mathcal{N}(-1,1)\,T[-\infty,0]$\tablenotemark{a} & Slope of SNeIa DTD & $\Psi = t^{-\alpha}$ for $t>t_{\mathrm{min}}$; $\Psi=0$ for $t<t_{\mathrm{min}}$ \\
$t_{\mathrm{min}}$ & $\Gamma(1,0.5)$ & Minimum delay-time of SNeIa & --- \\
\hline
$\eta$ & $\mathrm{LogNormal}(0.75,50)$ & Mass-loading factor & $\dot{M}_{\mathrm{out}} = \eta \dot{M}_{\star}$ \\
\hline
$f$ & $U(0,0.25)$ & Mixture weight & --- \\
\enddata
\tablenotetext{a}{The notation $T[-\infty,0]$ denotes truncation on the interval $(-\infty,0)$.}
\end{deluxetable*}

\subsection{Constructing the AZR} \label{sec:azr}

\begin{figure}
    \centering
    \includegraphics[width=\columnwidth]{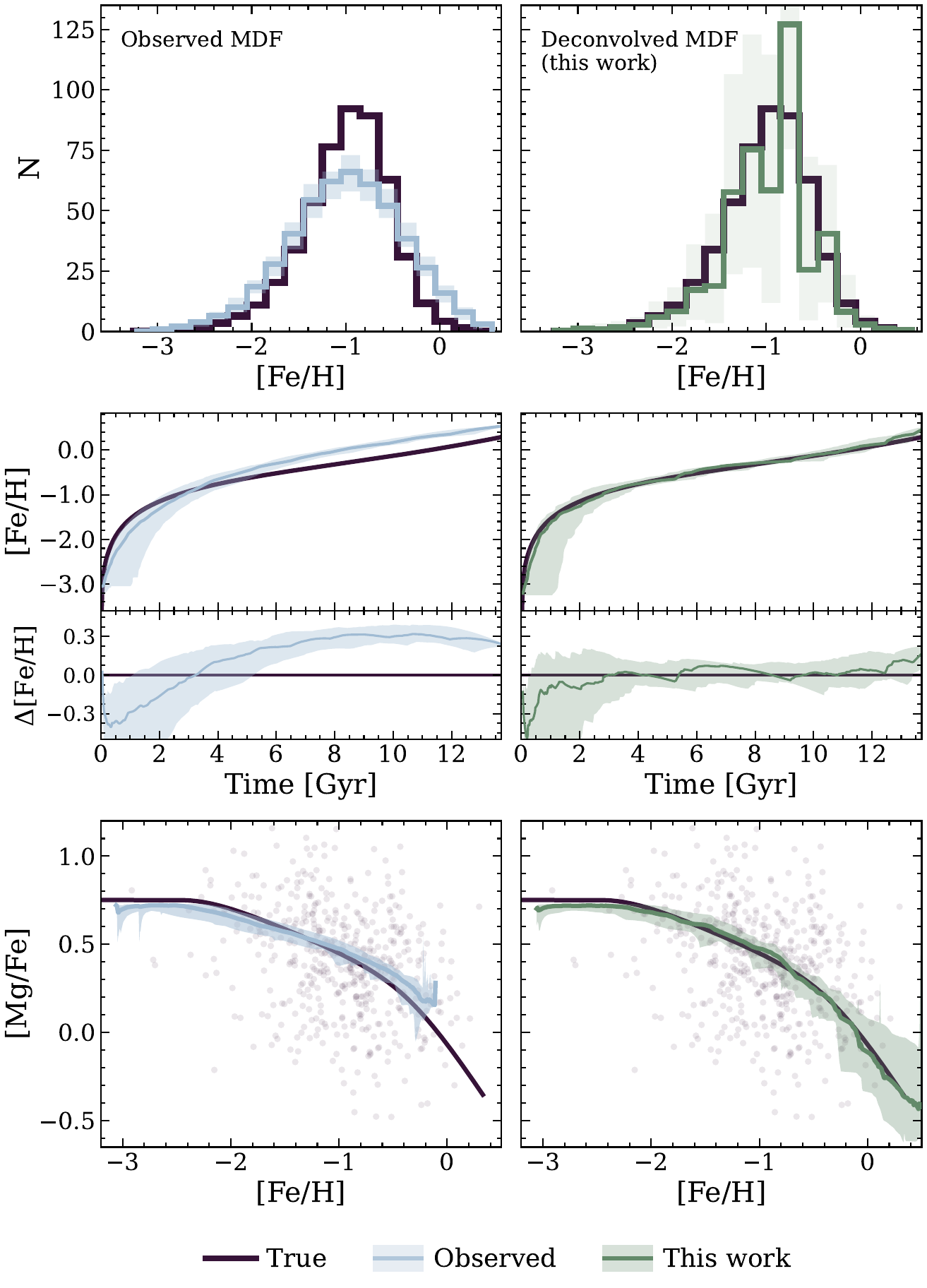}
    \caption{The downstream effects of using the observed MDF (blue) versus the deconvolved MDF (green) in DLEIY. The observed MDF (top row) leads to more significant bias in the AZR (middle row), which induces bias in the predicted chemical abundance track (bottom row).}
    \label{fig:azr_decon}
\end{figure}

Uncertainty in the MDF is handled differently than other sources of uncertainty.
Instead of using Monte Carlo draws directly from the observed MDF, we first deconvolve the measurement uncertainty using a hierarchical Bayesian model, and use posterior samples.

This step is necessary because measurement uncertainty in the MDF biases the AZR, as illustrated in Figure \ref{fig:azr_decon}.
Even when the uncertainties are homoskedastic and symmetric, their effect on the cumulative distribution function (CDF) is not.
As the MDF is broadened, the tails become more populated, the range increases, and the peak is largely unchanged, which corresponds to a flatter, wider CDF.
A quantile below the median will be reached at a lower metallicity, because the minimum has decreased and the number of stars in the tails has increased.
A quantile above the median will be reached at a higher metallicity.
The AZR will therefore underestimate the true metallicity at earlier times and overestimate at later times, as shown in the middle panel of Figure \ref{fig:azr_decon}.

Poor sampling of the tails due to a finite sample size can also bias the AZR; in particular, the minimum metallicity will tend to be overestimated (although this is mitigated somewhat by random uncertainties tending to increase the range).
This systematically shifts the AZR forward in time, because the minimum/maximum metallicities are anchored by the SFH.

These biases manifest in the predicted [$\alpha$/Fe]-[Fe/H] tracks as a systematic shift in the $\alpha$ `knee' (the downturn associated with the onset of SNeIa) towards lower metallicities and a flatter `shin', as shown in the bottom panel of Figure~\ref{fig:azr_decon}. 
This will bias inference towards shorter delay-times and steeper slopes in the delay-time distribution.
Also, systematic overestimation of the metallicity has a first-order effect on the ejected mass (Equation \ref{eq:sinks}) but only a second-order effect on the source term (Equation \ref{eq:sources}), which can artificially exhaust the metal supply and truncate the [Mg/Fe]-[Fe/H] track.
Truncated tracks would be harshly penalized in the likelihood, which would lead to a preference for parameters that do not result in truncation, such as a low mass-loading factor.
In short, accounting for the biases in the observed MDF is extremely important to reduce bias in the parameter estimation.

Often, resampling techniques like bootstrapping or Monte Carlo experiments are used to minimize the effect of sample-to-sample variance on an estimator (such as the mean metallicity) or to quantify its uncertainty.
In this case, however, we are not concerned with a biased estimator; rather, we are concerned with accurately representing the tails of the distribution, which cannot be resolved with resampling.
To properly account for the effect of measurement uncertainties, we deconvolve them from the observed MDF using a hierarchical Bayesian model used in \cite{leistedt_hierarchical_2017}.
We elaborate on the method in Appendix \ref{sec:app_decon}.

\section{Validating DLEIY with mock galaxies}\label{sec:mocks}

\begin{figure*}
\begin{center}
\includegraphics[width=0.95\textwidth]{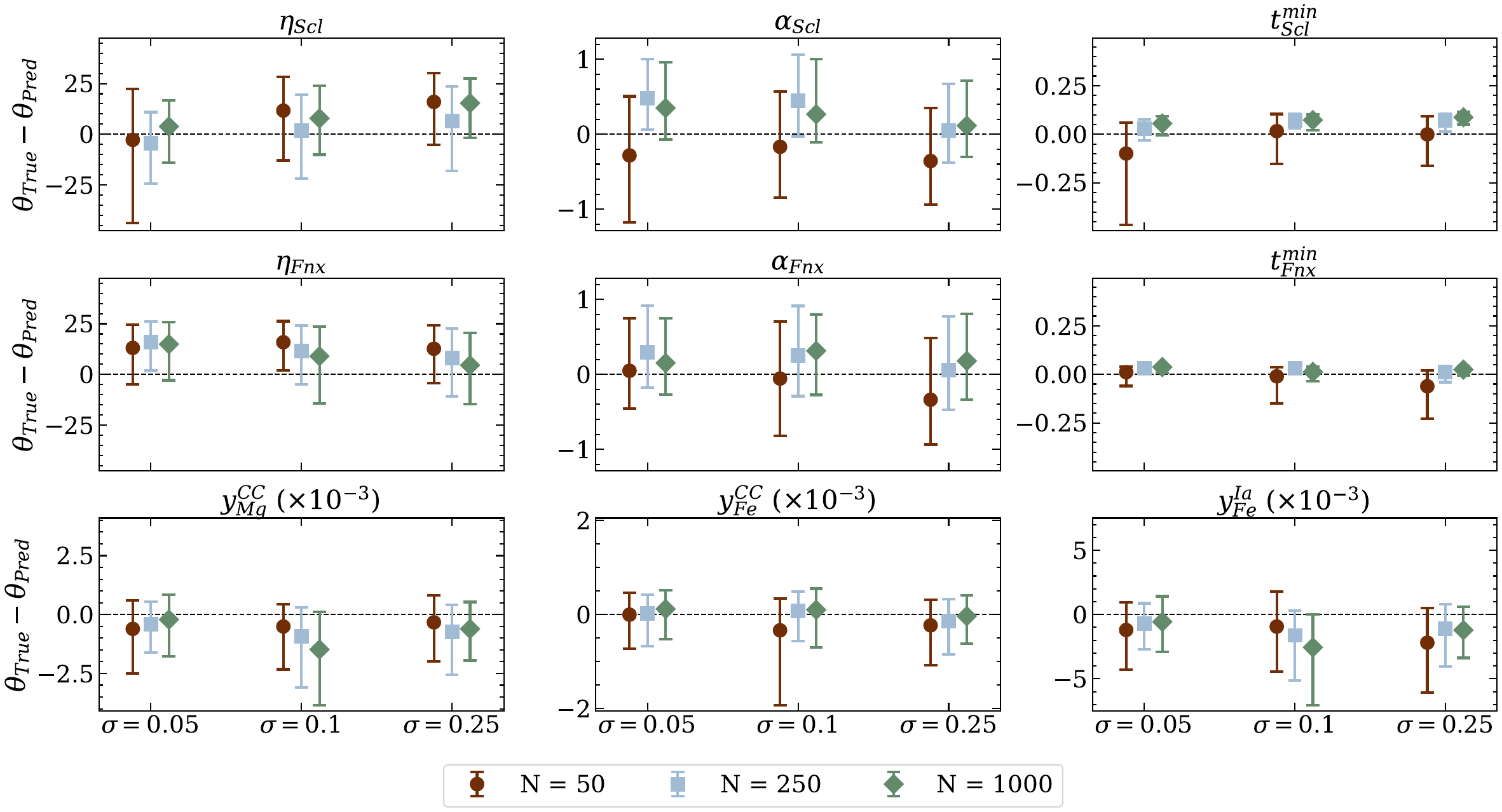}
\caption{Median residuals (with 16/84\% credible interval) for a grid of tests using mock galaxies made with VICE \citep{johnson_dwarf_2023}.
We use the shared-yield model on Sculptor- and Fornax-like mock galaxies for varying sample sizes and uncertainties. \label{fig:nsigma}}
\end{center}
\end{figure*}

To evaluate the reliability of DLEIY, we apply it to a series of mock galaxies constructed using the GCE model VICE \citep{johnson_dwarf_2023}.
We perform two suites of mock tests: one to evaluate the effect of sample size and measurement uncertainty, and one to demonstrate DLEIY's ability to discriminate between different DTDs.
For all tests, we use VICE in `infall mode', where the star formation rate is parameterized by the gas infall and a star formation efficiency, i.e., $\dot{M}_{g,~in}=I_{0}e^{-t/\tau_{in}}$ and $\tau_{\star}=M_{g}/\dot{M_{\star}}$
The parameters that describe the mocks are the infall normalization, infall timescale, star formation efficiency, initial gas mass, plus the same parameters as used in DLEIY ($\eta$, $y_{\mathrm{Mg}}^{CC}$, $y_{\mathrm{Fe}}^{\mathrm{CC}}$, $y_{\mathrm{Fe}}^{Ia}$, $\alpha$, $t_{\mathrm{min}}$).
See Appendix \ref{sec:app_unc} for fiducial values of these parameters.

We construct fiducial Sculptor-like and Fornax-like mocks with the same yields for all tests.
The values of the other parameters are in Table \ref{tab:mocks}.
These mocks qualitatively reproduce the features of Sculptor (single short burst of star formation, broad MDF, easily visible plateau/knee feature) and Fornax (extended, almost constant star formation, narrow MDF), as well as their masses and mean metallicities \citep{weisz_star_2014, kirby_multi-element_2010}.
We also include an outlier-generating process to mimic contamination, where we replace 2.5\% of the total sample with draws from a uniform distribution with bounds set by the extrema of the data $+0.2$ dex. 
When generating observed samples, we use the same deconvolution method that is applied to observed data.

For the first set of tests, visualized in Figure~\ref{fig:nsigma}, we run DLEIY with the shared-yield model for a grid of different sample sizes $N=(50,250,1000)$ and measurement uncertainties on the abundances $\sigma_{\mathrm{[Fe/H]}}=\sigma_{\mathrm{[Mg/Fe]}}=(0.05, 0.1, 0.25)~\textrm{dex}$.
A new sample is drawn for each model; we do not use the same 50-star sample for all uncertainties, for example.
As shown in Figure \ref{fig:nsigma}, the constraints from a small sample size ($N=50$) can be highly variable, even when the uncertainty is small.
We attribute this variance to the fact that stars in the metal-poor tail provide the strongest constraints, but the tail is poorly sampled with only $N=50$ stars, and small observational uncertainties cannot overcome this deficit.
That the constraint for some parameters does not improve with sample size or uncertainty suggests that the uncertainty in the SFH dominates.

There is some bias to slightly high $y_{\mathrm{Mg}}^{CC}$ and $y_{\mathrm{Fe}}^{Ia}$ yields and low $\eta_{\mathrm{Fnx}}$, although the uncertainties are appropriate.
These three parameters have log-normal prior distributions, so a slight bias is not surprising.
The two yields are also correlated on approximately the same scale, and the offsets are comparable, as expected.
We did test uniform priors with real data, but sampling was less well-behaved with priors with finite support, so we did not run the mock tests using uniform priors.
Uniform priors would be expected to reduce the bias, but worsen the uncertainty (as they are less informative).

To evaluate whether DLEIY can be used to reliably discriminate between different DTDs, we vary the slope to be either $-1$ or $-2$, with a minimum delay of 0.05 Gyr for the Fornax-like mock and 0.15 Gyr for the Sculptor-like mock.
The sample size is the same as the actual observed sample size, and we approximate the observed uncertainties in the SFH and abundances, as detailed in Appendix \ref{sec:app_unc}.

\begin{figure}
\begin{center}
\includegraphics[width=0.9\columnwidth]{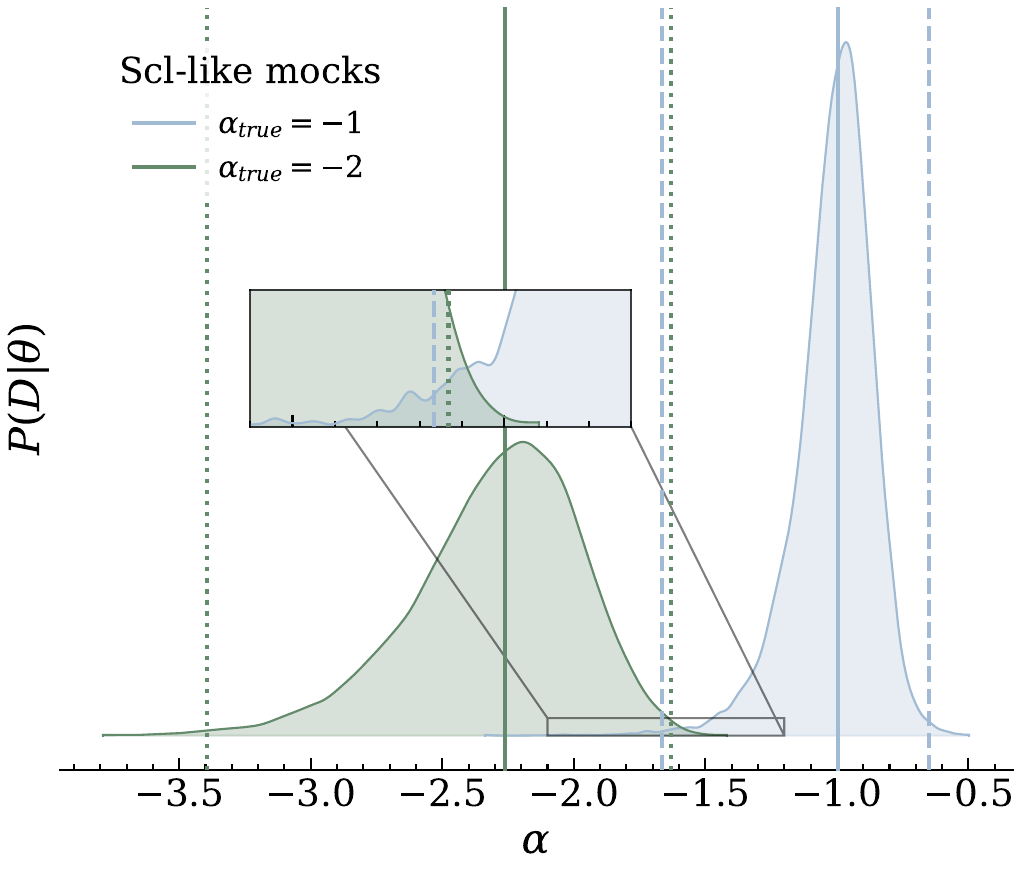}
\caption{Posterior distributions for the DTD slope parameter $\alpha$ for two mock Sculptor-like galaxies with different DTDs, with the median and 0.3/99.7\% credible interval demarcated by vertical lines. The region of overlap is highlighted. The distributions can be distinguished at $\sim2.5\sigma$. This mock test was run using the shared-yield model.
\label{fig:mock_slopes}}
\end{center}
\end{figure}

The primary result of this test is demonstrated in Figure \ref{fig:mock_slopes}, which shows the posterior distribution of the slope parameter $\alpha$ for a Sculptor-like mock with true slope $\alpha=-1$ or slope $\alpha=-2$, with dashed/dotted lines showing the 0.3/99.7\% credible interval (approximately 3$\sigma$).
The parameters can be distinguished at $\sim2.5\sigma$.
In this test, the slopes for the Fornax-like mock were both $\alpha=-2$.

The Fornax-like mock is less sensitive to the slope, and this is reflected in larger uncertainties.
We touch on this further in Section \ref{sec:DTD}.
For a true slope of $\alpha=-2$, the median and 0.3/99.7\% credible interval were $\alpha=-2.26^{+1.75}_{-2.04}$.
For a true slope of $\alpha=-1$, the median and 0.3/99.7\% credible interval were $\alpha=-1.96^{+0.97}_{-2.03}$, so it is biased but consistent with the truth within $\sim3\sigma$.
In this test, the slopes for the Sculptor-like mock were both $\alpha=-2$.

\section{Modelling Sculptor and Fornax}\label{sec:apply}

\begin{figure*}
\begin{centering}
\includegraphics[width=\textwidth]{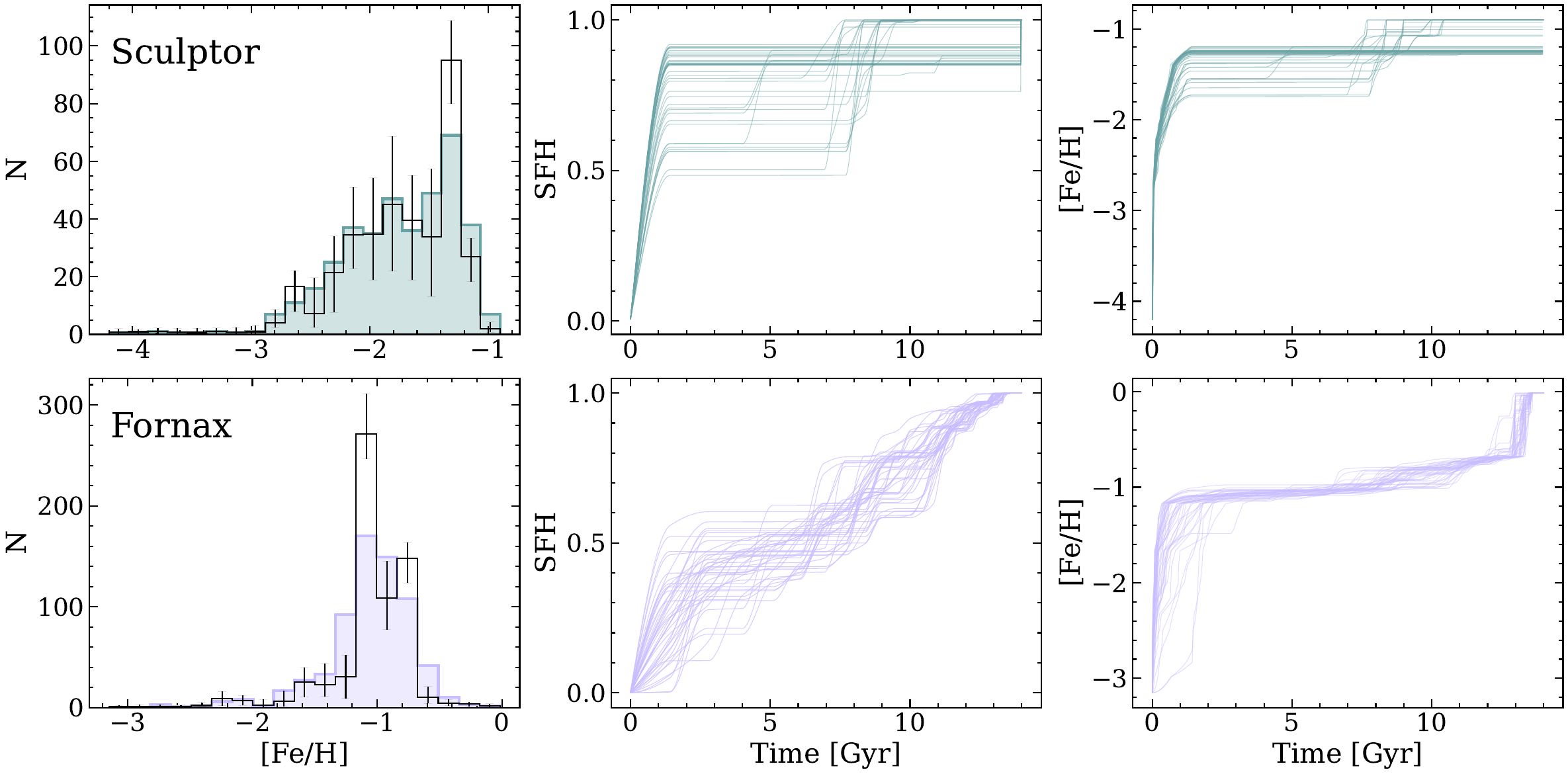}
\caption{(Left) Observed MDFs of Sculptor (top) and Fornax (bottom) \citep{kirby_multi-element_2010,henderson_neutron-capture_2025}, overlaid with the median de-convolved MDF and 16/84\% credible intervals in black. (Middle) Monte Carlo draws from the observed SFHs \citep{weisz_star_2014}. (Right) Ensemble of age-metallicity relations made using the Monte Carlo draws of the SFH and the de-convolved MDF. \label{fig:dat}}
\end{centering}
\end{figure*}
 
We now apply DLEIY to real data from the dSphs Sculptor and Fornax, using abundances from \cite{kirby_multi-element_2010} and \cite{henderson_neutron-capture_2025}.
We use the updated measurements from \cite{henderson_neutron-capture_2025} whenever available, because the Mg abundance is measured from stronger features (the Mg b triplet), resulting in a more reliable abundance.
The SFHs are based on the work of \cite{weisz_star_2014}, but we recalculated the \cite{weisz_star_2014} solutions rather than using the publicly available SFHs. 
This is necessary because the public SFHs only provide the fiducial solution and the uncertainty intervals, whereas our analysis requires access to the full set of random SFH realizations used in the uncertainty derivation.

Briefly, we calculate our SFHs using the public photometry and artificial star tests from \cite{weisz_star_2014}, for a photometric field in Sculptor (field u43502) and Fornax (field u2lb02). The SFHs have been calculated to mimic the \cite{weisz_star_2014} methodology as closely as possible. We use the same software for the color-magnitude diagram (CMD) fitting \citep[MATCH;][]{dolphin_numerical_2002}, stellar models, distance and extinction values, binary fraction, initial mass function, and metallicity treatment as \cite{weisz_star_2014}. 
Due to minor differences in the CMD-fitting scheme (i.e., age binning, CMD binning, photometric depth), and to the use of an updated version of MATCH, the SFHs we obtain are not perfectly identical to the \cite{weisz_star_2014} solutions, but the differences are minor and entirely compatible with the associated uncertainties.

The different SFH draws are obtained following the procedure outlined in \cite{dolphin_estimation_2012} for the derivation of systematic uncertainties. 
In brief, random perturbations are applied to the luminosity and $T_{eff}$ scales of the select stellar models, which are then used to calculate a new SFH solution. 
The stellar model perturbations are calibrated to roughly reproduce observed differences between different stellar model libraries. 
This procedure is repeated 100 times and the resulting SFH set is used as the basis for our uncertainty exploration.

The observed and deconvolved MDF, SFH draws, and AZRs are presented in Figure \ref{fig:dat}. 
We chose Sculptor and Fornax because they are well-studied systems with a large amount of data that have very different SFHs.
As shown in Figure \ref{fig:dat}, Sculptor forms most of its stars in a single burst, within a couple of Gyr, while Fornax has nearly constant star formation for nearly the age of the universe.
Their masses are also different, which is important to breaking the yield/outflow degeneracy, as the mass-loading factor is expected to be mass-dependent \citep{murray_maximum_2005}.

\begin{figure*}
\begin{centering}
\includegraphics[width=\textwidth]{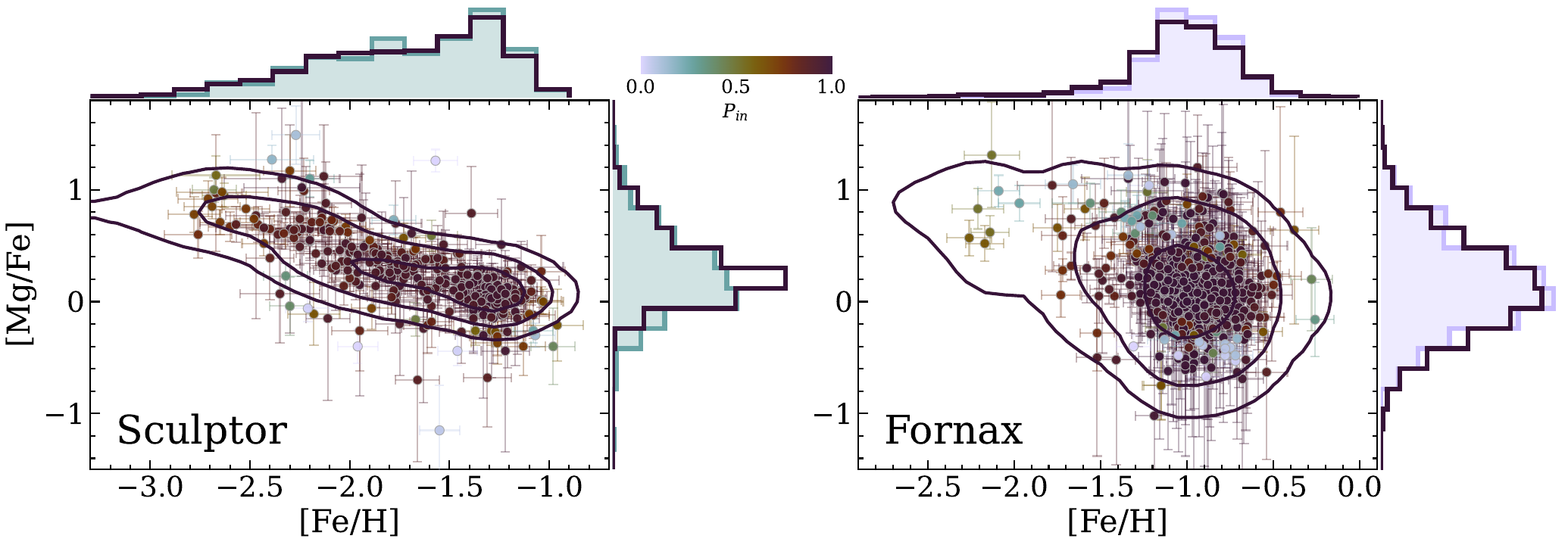}
\caption{Posterior predictive plots for Sculptor (Left) and Fornax (Right). The 40/86/97.5\% credible intervals of the predicted distribution are plotted as purple contours, with the observed data (colored by their average posterior probability of being an inlier) over-plotted. The filled histograms show the marginal distributions of the observed data, and the purple histograms show the predicted distribution. The posterior predictive is convolved with uncertainty, but does not include the outlier generating process.
\label{fig:postpred}}
\end{centering}
\end{figure*}

The parameter estimates and uncertainties are presented in Table \ref{tab:results}.
Figure \ref{fig:postpred} shows the posterior predictive plot, which compares the observed data to data generated with draws from the posterior, convolved with a measurement uncertainty. 
The predicted [Mg/Fe] distribution for Sculptor is a good match for the mean and the higher [Mg/Fe] tail, although it has a slightly stronger peak and less prominent tail to low [Mg/Fe].
The predicted [Mg/Fe] distribution for Fornax is slightly over-dispersed around the mean, but the tails are captured well and the location of the mean is a good qualitative match for the true distribution.
We also show the probability that a point is an inlier, with a prior that $3\%$ ($1.3\%$) of the points in Sculptor (Fornax) are outliers (based on the number of points identified as outliers using the local outlier factor).
Because DLEIY has no intrinsic scatter, points with small uncertainties that are far from most predicted model tracks are more likely to be considered outliers, resulting in the areas of low probability around the centroid for Fornax.

The independent and shared-parameter models are all highly consistent with each other.
We compare the models using the evidence by calculating the Bayes factor: $K=\mathcal{Z}_{1}/\mathcal{Z}_{2}=P(D|M_1)/P(D|M_2)$. 
This is similar to a likelihood-ratio test.
When comparing the shared-yield and shared-yield+DTD model, there is a weak preference for the shared-yield+DTD model ($\log_{10}{K}=-1.1$).
The independent models are strongly preferred to the shared-parameter models, although we note that the independent models were run separately (sequentially) for computational reasons, so we can only compare the independent model of Sculptor or the independent model of Fornax to a shared-parameter model, which is not necessarily a fair comparison.
Specifically, the Bayes factor of the shared-yield+DTD (shared-yield) versus independent model of Sculptor is $\log_{10}{K}=-381.7$ ($\log_{10}{K}=-382.8$), and versus the independent model of Fornax is $\log_{10}{K}=-275.5$ ($\log_{10}{K}=-276.6$).

\def\arraystretch{1.15}
\begin{deluxetable}{lcccc}
\tablecaption{Parameter estimates and 16/84\% credible intervals for the shared-yield, shared-yield+DTD, and independent models of Sculptor and Fornax. $t_{\mathrm{min}}$ has units of Gyr; other parameters are unitless. \label{tab:results}}
\tablehead{
\\
\colhead{$\theta$} &
\colhead{\parbox[c]{2.0cm}{\centering Shared--\\yield}} &
\colhead{\parbox[c]{2.0cm}{\centering Shared--\\yield+DTD}} & 
  \colhead{\parbox[c]{1.6cm}{\centering Scl\\(indep)}} &
  \colhead{\parbox[c]{1.6cm}{\centering Fnx\\(indep)}}\\
}
\startdata
$\eta_{\mathrm{Scl}}$ & $ 46.3^{+29.0}_{-21.2}$ & $ 42.1^{+29.3}_{-18.5} $& $ 37.6^{+29.3}_{-17.7}$ & \nodata \\
$\eta_{\mathrm{Fnx}}$ & $ 34.8^{+17.8}_{-14.2} $& $ 30.0^{+18.9}_{-13.1} $& \nodata & $  33.6^{+18.9}_{-14.4} $ \\ \midrule
$y_{\mathrm{Mg}}^{CC}$ & $ 3.81^{+1.23}_{-1.02} $& $ 4.20^{+1.85}_{-1.26} $& $ 3.62^{+1.53}_{-1.05}$ & $  3.88^{+1.65}_{-1.28} $ \\
$y_{\mathrm{Fe}}^{CC}$ & $ 1.18^{+0.53}_{-0.38} $& $ 1.14^{+0.54}_{-0.38} $& $ 1.12^{+0.53}_{-0.40}$ & $  1.15^{+0.84}_{-0.46} $ \\
$y_{\mathrm{Fe}}^{Ia}$ & $ 4.61^{+1.92}_{-1.48} $& $ 5.28^{+2.97}_{-1.80} $& $ 4.79^{+2.57}_{-1.67}$ & $  4.51^{+2.81}_{-1.98} $ \\ \midrule
$\alpha_{\mathrm{Scl}}$ & $-2.24^{+0.48}_{-0.61} $& $-2.13^{+0.45}_{-0.61} $& $-2.10^{+0.52}_{-0.64}$ & \nodata \\
$\alpha_{\mathrm{Fnx}}$ & $-1.98^{+0.48}_{-0.66} $& $-2.13^{+0.45}_{-0.61} $& \nodata & $ -1.98^{+0.53}_{-0.74} $ \\
$t^{\mathrm{min}}_{\mathrm{Scl}}$ & $ 0.14^{+0.05}_{-0.05} $& $ 0.12^{+0.06}_{-0.04} $& $ 0.14^{+0.06}_{-0.05}$ & \nodata \\
$t^{\mathrm{min}}_{\mathrm{Fnx}}$ & $ 0.05^{+0.07}_{-0.03} $& $ 0.12^{+0.06}_{-0.04} $& \nodata & $  0.05^{+0.07}_{-0.04} $  \\
\enddata
\end{deluxetable}

\subsection{Mass-loading factor}
\begin{figure}
\begin{centering}
\includegraphics[width=\columnwidth]{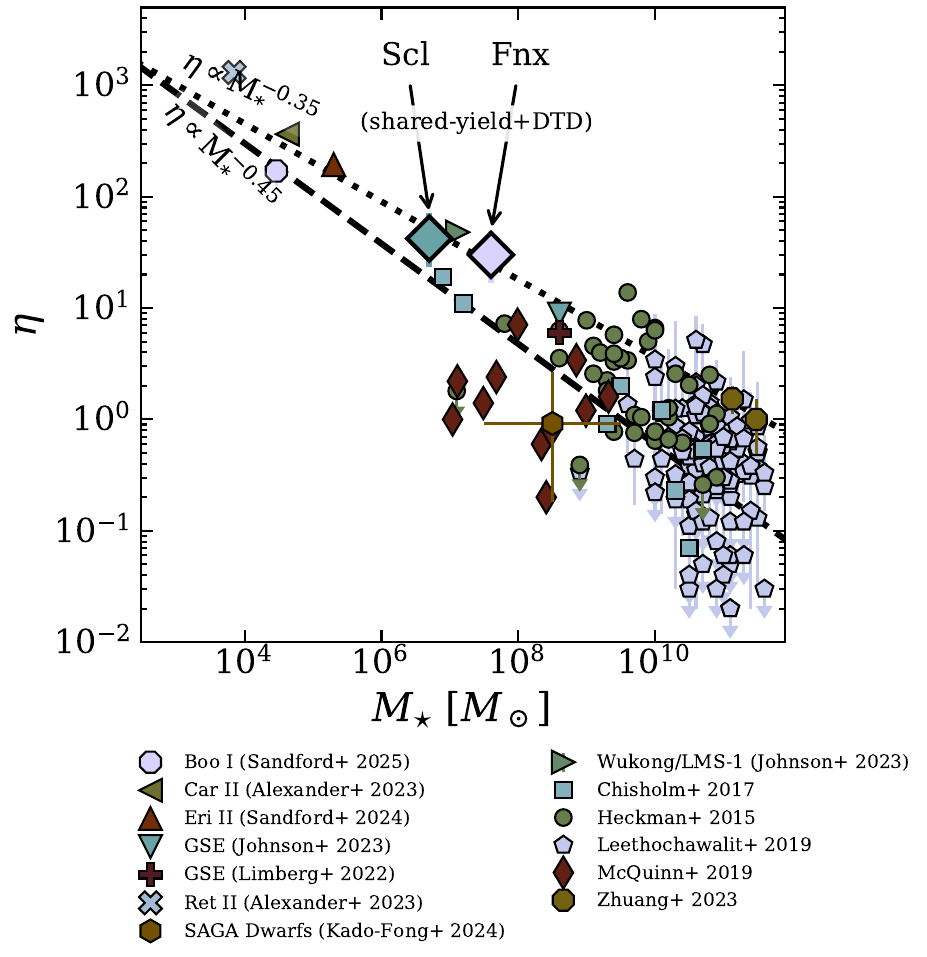}
\caption{Mass-loading factor $\eta$ and the 16/84\% credible interval for Sculptor and Fornax from the shared-yield+DTD model, compared to mass-loading factors in literature \citep{heckman_systematic_2015, chisholm_mass_2017,leethochawalit_evolution_2019,mcquinn_galactic_2019,limberg_reconstructing_2022,alexander_inhomogeneous_2023,johnson_dwarf_2023, zhuang_glimpse_2023,kado-fong_sagabg_2024,sandford_strong_2024,sandford_chemodynamics_2025}.
The dashed line shows the typical scaling from the FIRE-2 simulations, which is indicative of energy-driven winds \citep{pandya_characterizing_2021}.
The dotted line shows the typical scaling from FIRE-1, which is indicative of momentum-driven winds \citep{muratov_gusty_2015}.
Both Sculptor and Fornax fall along the general trend of the observed data.
\label{fig:eta}}
\end{centering}
\end{figure}

The mass-loading factors, shown in Figure \ref{fig:eta}, align with expectation: as the potential well deepens, the strength of outflows decreases.
The values from DLEIY are consistent with the trend of other measurements from GCE models \citep[e.g.,][]{sandford_strong_2024} and many of the direct observations \citep[e.g.,][]{heckman_systematic_2015, chisholm_mass_2017}.
There is some tension with observations that measure lower mass-loading factors or only a weak mass dependence \citep[e.g.,][]{mcquinn_galactic_2019}.
The mass-loading factors from this work also align with the scaling relation and normalization from the FIRE-1 simulations, which are indicative of momentum-driven winds.
Mass-loading factors inferred from other GCE models also tend to agree with this higher normalization \citep[e.g.,][]{alexander_inhomogeneous_2023,johnson_dwarf_2023,sandford_strong_2024,sandford_chemodynamics_2025}.
We emphasize that, unlike other measurements from GCE models, these measurements are not relative to the nucleosynthetic yields; they are determined jointly with the absolute scale of yields.
Given the uncertainties in measurements of $\eta$ and the scatter in this relationship, the steeper scaling characteristic of energy-driven winds from the FIRE-2 simulations cannot be ruled out.

\subsection{Delay-time distribution}
The minimum delay for SNeIa aligns with expectations, ranging from $\sim0.05-0.15$ Gyr \citep{maoz_star_2017,poulhazan_precision_2018}.
In the independent and shared-yield model of Fornax, a short minimum delay is preferred, while in all models of Sculptor, a longer minimum delay is preferred, although they are consistent with each other within $1\sigma$. 

With all models, we find a steep DTD of $\sim t^{-2}$, in contrast with the $t^{-1}$ found by most transient surveys \citep[e.g.,][]{maoz_type-ia_2012}. 
The measurements of $\alpha$ from all models are inconsistent with $-1$ within the 16/84\% credible interval.
All but the independent model of Fornax are inconsistent within a more stringent 0.3/99.7\% credible interval (about $3\sigma$). 

In general, Sculptor-like galaxies are more sensitive to the DTD, because galaxies with extended star formation like Fornax tend to reach an equilibrium abundance quickly relative to the duration of star formation.
As a result, most stars form at equilibrium, well after the onset of SNeIa, so the constraint especially on the minimum delay is weaker.
While abundances are sensitive to very short timescales, the star formation histories affect the precision of the constraints on the slope and especially the minimum delay.

\subsection{Yield-outflow degeneracy}
In many GCE models, a low-yield, low-outflow galaxy will mimic a high-yield, high-outflow galaxy.
This yield-outflow degeneracy is such that a change in yield will induce a change in the outflow on a similar order of magnitude (and vice-versa) \citep{johnson_dwarf_2023}.
DLEIY's unique approach to jointly modeling multiple galaxies at once can help break this degeneracy.

\begin{figure*}
\begin{centering}
\includegraphics[width=0.8\textwidth]{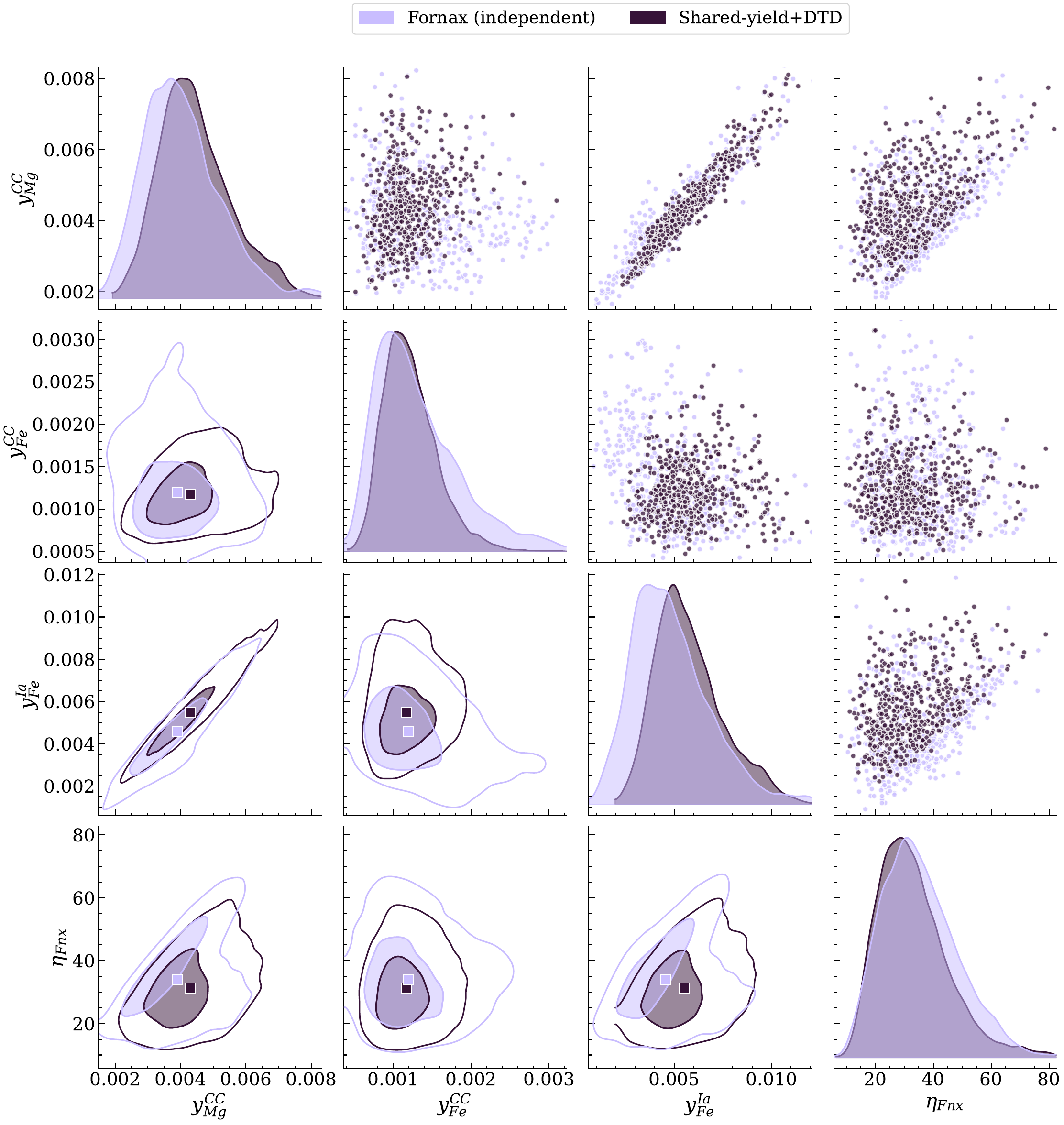}
\caption{Pair plot highlighting the yield/outflow degeneracy, with the 40\% and 84\% contours (approximately 1 and 2$\sigma$) on the lower diagonal, with the medians marked as squares, and 500 posterior samples on the upper diagonal.  Lavender is the independent model of Fornax only and purple is the shared-yield+DTD model. The bottom left plot most clearly demonstrates how the yield-outflow degeneracy is much weaker in the shared-yield+DTD model than the independent model for Fornax.
\label{fig:yieldoutflow}}
\end{centering}
\end{figure*}

Figure \ref{fig:yieldoutflow} shows a corner plot highlighting this effect.
The bottom row most clearly illustrates how sharing yields between galaxies improves the yield-outflow degeneracy.
In the independent model of Fornax, $y_{\mathrm{Mg}}^{CC}$ and $y_{\mathrm{Fe}}^{Ia}$ are strongly correlated with $\eta_{\mathrm{Fnx}}$.
While the point estimates are consistent between all models, the joint model significantly reduces the degree of correlation and decreases the uncertainty.
The correlation between $y_{\mathrm{Mg}}^{CC}$ and $y_{\mathrm{Fe}}^{Ia}$, which arises in part because the ratio of these parameters has a strong effect on the equilibrium abundance, does not significantly resolve in the shared-parameter models.
See Figure \ref{fig:corner_all} for the full corner plot of all parameters and models.

%

\section{Discussion}\label{sec:discussion}
\subsection{Yields}\label{sec:yields}

\begin{figure*}
\begin{center}
\includegraphics[width=0.95\textwidth]{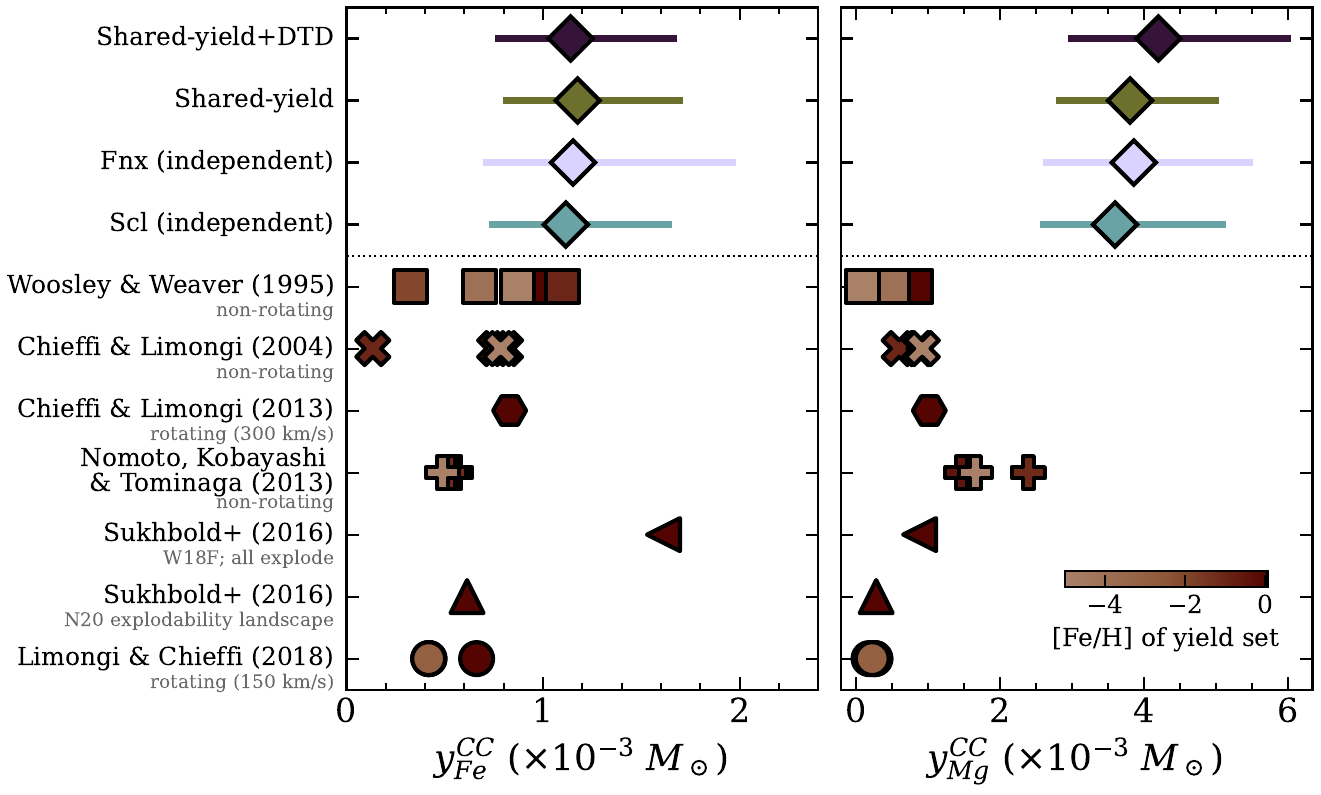}
\caption{Comparison of population-averaged Fe and Mg yields for DLEIY (diamonds) and an integrated yield from yield tables in literature \citep{woosley_evolution_1995,chieffi_explosive_2004, chieffi_pre-supernova_2013,nomoto_nucleosynthesis_2013, sukhbold_core-collapse_2016, griffith_impact_2021, limongi_presupernova_2018}, assuming a \citet{kroupa_variation_2001} initial mass function. The error bars on the DLEIY measurement are the 16/84\% credible interval. \label{fig:cc_yields}}
\end{center}
\end{figure*}

The choice of stellar yields is a well-documented source of significant uncertainty in GCE models \citep[e.g.,][]{romano_quantifying_2010, molla_galactic_2015, cote_impact_2017, poulhazan_precision_2018,liang_assessing_2023}.
Stellar yield models are subject to uncertainties in nuclear physics and stellar physics, and require simplifying assumptions about rotation, symmetry, mass loss, etc. 
Explodability prescriptions, gridding in mass and metallicity, and the range of masses and metallicities in the yield set can also result in differences in the population-averaged yield in excess of 0.5 dex \citep{cote_mass_2016}.

As mentioned, yields are often not the parameters of interest in a GCE model, so they will be fixed according to a yield set or tuned manually to reproduce a particular feature of a dataset like the $[\alpha/\textrm{Fe}]$ plateau \citep[e.g.,][]{schonrich_chemical_2009,vincenzo_chemical_2014,dubay_galactic_2024}.
However, the choice of yields \emph{does} affect the inference of those parameters \citep{philcox_inferring_2019}---the yield-outflow degeneracy (Figure~\ref{fig:yieldoutflow}) is a clear example of this.
There are also degeneracies between yields, the star formation efficiency, the delay-time distribution, and possibly other parameters depending on the parameterization of the model \citep{andrews_inflow_2017}.
In general, the absolute scale of yields has implications for the inference and understanding of other parameters that govern galaxy evolution.
For example, the absolute scale of outflows and yields are relevant to reproducing the observed mass-metallicity relation and the enrichment of the circumgalactic medium \citep{peeples_budget_2014, weinberg_scale_2024}.
Empirical yields also represent an independent constraint on nucleosynthetic processes that can inform supernova yield models.
As such, the choice of yield set is not benign, and can indeed be a source not just of uncertainty, but of bias.

Here, we compare the measurements of the population-averaged yields from DLEIY and yield tables in literature.
Measurements from DLEIY and yield tables are not directly comparable, which is worth keeping in mind when interpreting the results. 
In DLEIY, we measure a population-averaged yield directly.
The yield tables report the yield as a function of progenitor mass and metallicity, which we linearly interpolate and integrate (with some assumptions about explodability) over a \citet{kroupa_variation_2001} initial mass function.
A discrepancy between the measured population-averaged yield and this integrated yield could be due to any of these factors, and it is difficult to know which is the most important. 

\subsubsection{CCSN Fe yields}

The left panel of Figure \ref{fig:cc_yields} shows a comparison of the population-averaged Fe CCSN yields from DLEIY and the integrated yield from various yield tables in the literature \citep{woosley_evolution_1995,chieffi_explosive_2004, chieffi_pre-supernova_2013, nomoto_nucleosynthesis_2013, sukhbold_core-collapse_2016, griffith_impact_2021, limongi_presupernova_2018}.
The yield tables have a diverse set of assumptions about metallicity, mass range, and rotation, and the integrated yield varies between yield tables by about a factor of 2.

The `explodability' landscape also varies for different yield tables.
Previous work has suggested that rather than a mass range in which all stars explode, above which they directly collapse, there are instead `islands of explodability', relatively narrow mass ranges in which stars tend to explode and others where they tend to collapse \citep{oconnor_black_2011,ugliano_progenitor-explosion_2012}.
The \sk~models with the N20 engine have a self-consistent explodability landscape (see their Figure 13); the W18F models are an adaptation of the W18 engine in which all progenitors explode, implemented in \cite{griffith_impact_2021}.
The preferred explodability with the \lc~models is that all models between $13-25M_{\odot}$ explode, and more massive stars collapse directly.
For \ww, \clof, and \cltt, all stars up to some upper mass limit are assumed to explode ($40 M_{\odot}$, $35 M_{\odot}$, and $40M_{\odot}$ respectively). 
We comment on this variety to emphasize that two models' consistency in integrated yield can hide discrepant explodability and/or ejected mass as a function of progenitor mass or metallicity.  

Compared to these theoretical integrated yields, we find that a slightly higher yield is preferred, but this is not highly statistically significant. 
The 16/84\% credible intervals from DLEIY are consistent with many of the theoretical models (Figure \ref{fig:cc_yields}).
A few of the \ww~and \clof~models are excluded at a higher confidence ($0.3/99.7\%$).

Realistically, a stronger constraint is needed to make a meaningful distinction between theoretical yields.
The slight improvement in the constraint from the shared-parameter models highlights that a larger sample of galaxies would likely improve the constraint.
Theoretically, we do expect some galaxies to be more sensitive to the yields than others.
For example, a galaxy with a well-defined plateau (e.g., due to high star formation efficiency) and a short duration of star formation may provide stronger constraints.
The most constraining power is expected to come from the more metal-poor stars, so exploiting a metallicity gradient and using a sample that includes stars at larger radii could also improve the estimate, although this could worsen any mismatch between the populations described by the SFH and MDF.
Galaxies that reach a chemical equilibrium may also be more constraining.
We refer to \cite{weinberg_equilibrium_2017} for an extensive discussion of this phenomenon, but briefly, the [Mg/Fe] abundance typically approaches an equilibrium when there is continuing enrichment from both CCSNe and SNeIa (usually due to ongoing star formation), while the [X/H] abundances reach an equilibrium if the production of new metals and the dilution and depletion of existing metals are balanced.
The equilibrium [Mg/Fe] abundance depends most strongly on the Fe and Mg yields, with a lesser dependence on the delay-time distribution, outflows, or star formation history \citep{weinberg_equilibrium_2017}.


\subsubsection{CCSN Mg yields}

As shown in the right panel of Figure~\ref{fig:cc_yields}, the measured Mg yields from DLEIY are higher than the theoretical integrated yields, with the median being about a factor of 4 higher than the typical integrated yield.
This discrepancy is statistically significant.
The measurements from DLEIY are discrepant with the integrated yield from all yield tables within the 16/84\% credible interval, and discrepant with all but the integrated \nkt~yields within the 0.01/99.9\% credible interval (about $5\sigma$).

This discrepancy is not especially surprising, as underproduction of Mg in multiple yield sets has been previously identified in literature \citep[e.g.,][]{philcox_optimal_2018}.
For example, \cite{timmes_galactic_1995} demonstrated that [Mg/Fe] is underestimated in their GCE model of the Milky Way using yields from \ww.
Adjusting the Fe yield improves this discrepancy around solar metallicities, but [Mg/Fe] is still underestimated for metal-poor stars, suggesting it is indeed underproduced.
\cite{andrews_inflow_2017} found a similar result with the \ww~yields and \clof~yields. 
Also, the yields from \lc~and \sk~have [Mg/O] $< 0$, suggesting that Mg is underproduced and/or O is overproduced; \cite{griffith_impact_2021} suggest that likely both are true.

The constraint from DLEIY provides a more model-independent confirmation of results from these previous GCE and nucleosynthesis studies, as no assumptions are made about the other yields, equilibrium abundance, explodability, etc. 
There are many possible sources of this discrepancy.
A low triple-$\alpha$ or high $^{12}$C$(\alpha,\gamma)$ reaction rates in theoretical yield models could depress the Mg abundance \citep{griffith_impact_2021}, as these reactions affect the O abundance and the availability of $^{12}$C seeds used to produce $^{24}$Mg. 
The mass loss prescription in these models can also affect yields of lighter elements without significantly changing the Fe yield, and stellar rotation or convection prescriptions affect the degree of internal mixing and therefore the availability of different nuclei for fusion \citep{griffith_impact_2021}.
Improved precision on this empirical measurement and the inclusion of more elements like O or Na, which is sensitive to the $^{12}$C$(\alpha,\gamma)$ rate \citep[e.g.,][]{tur_sensitivity_2007}, may help differentiate between these different scenarios.





\subsection{DTD and SNIa yield}\label{sec:DTD}

\begin{figure*}
\begin{center}
\includegraphics[width=0.95\textwidth]{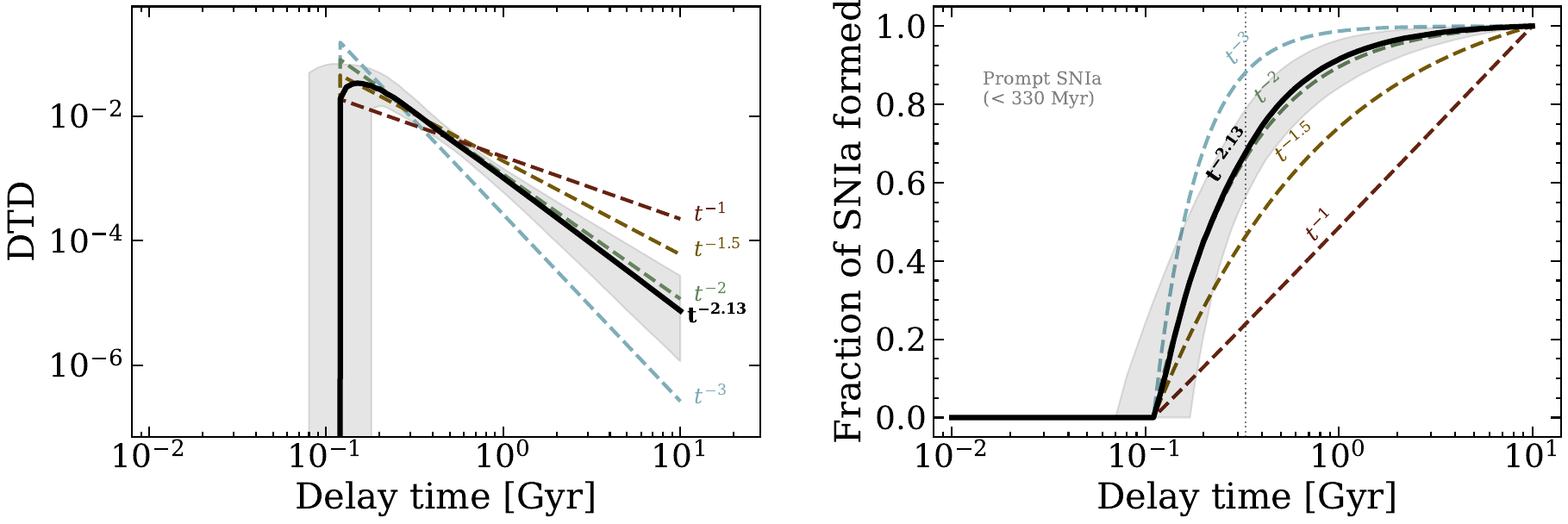}
\caption{(Left) Comparison of a power-law DTD with varying slopes. The distribution from the shared-yield+DTD model is in black, with the 16/84\% credible interval in gray. (Right) Cumulative distribution of SNeIa for power-law DTD with varying slopes. The dotted line approximately demarcates `prompt' versus `tardy' SNeIa. A steeper power-law slope corresponds to a larger fraction of prompt SNeIa and more rapid iron enrichment.
\label{fig:dtd}}
\end{center}
\end{figure*}

\begin{figure}
\begin{center}
\includegraphics[width=\columnwidth]{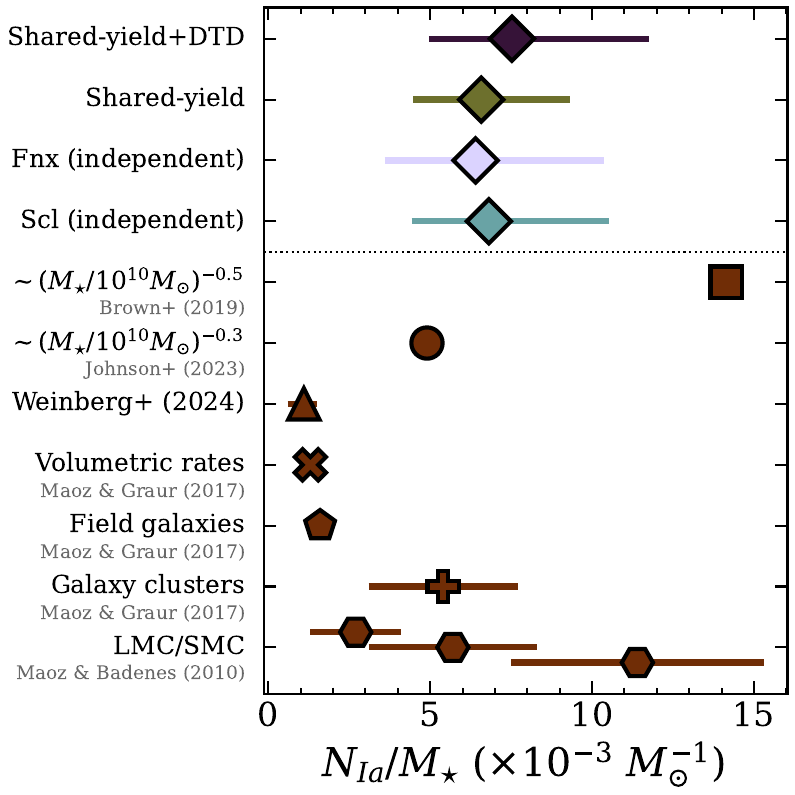}
\caption{Comparison of $N_{Ia}/M_{\star}$ for DLEIY (diamonds; 16/84\% credible interval shown), and measurements or scaling relations from transient studies \citep{maoz_supernova_2010,maoz_star_2017,brown_relative_2019} or GCE \citep{johnson_binaries_2023,weinberg_scale_2024}. We assume an average yield per SNIa of $m_{\mathrm{Fe}}^{Ia}=0.7M_{\odot}$, and for the scalings with mass, we use $M_{\star}=5\times10^{7}M_{\odot}$. We also note that the higher measurements for the LMC/SMC may have contamination from CCSNe and could be overestimated. \label{fig:nia}}
\end{center}
\end{figure}

With DLEIY, the inferred delay-time distribution of SNeIa for Sculptor and Fornax is a steep power-law of $t^{-2}$, with a minimum delay of $\sim0.05-0.15$ Gyr.
We emphasize that we use a power-law as a convenient analytic form.
SN surveys have generally converged to a power-law DTD of $t^{-1}$ \citep[e.g.,][]{maoz_supernova_2010}, although there is some observational evidence for a steeper delay in the literature \citep[e.g.,][]{heringer_delay_2019,chen_constraining_2021}.
This $t^{-1}$ shape is also a consequence of the double-degenerate progenitor channel, if the binary population has an initial distribution of orbital separations $P(a)\propto a^{-1}$ and a merger timescale $t\propto a^{4}$ \citep{ruiter_rates_2009}.
We do not wish to suggest that the measured power-law of $t^{-2}$ must have a similar origin, that it must be the consequence of one progenitor channel and a specific separation distribution. 
Instead, we focus on the implications of such a distribution.
Chiefly, this distribution suggests a significant fraction of prompt SNeIa (Figure \ref{fig:dtd}), with about 75\% of the SNeIa being formed by about 330 Myr after a burst of star formation.
In contrast, a $\sim t^{-1}$ distribution only forms about 25\% of the SNeIa by that time.
Iron enrichment after the onset of SNeIa is therefore very rapid.
Qualitatively, this manifests as a significant low-$\alpha$ peak in the [Mg/Fe] distribution and nearly absent high-$\alpha$ peak.

To compare the measured Fe yield from SNeIa to the results from SN surveys, we convert to a specific rate $N_{Ia}/M_{\star}$ (sometimes referred to as the SNIa efficiency).
The population-averaged yield that we infer is the product of the specific rate and the average yield per SNIa, so we adopt an average yield per SNIa to calculate the specific rate.
We set $m_{\mathrm{Fe}}^{Ia}=0.7M_{\odot}$ \citep[as in, e.g.,][]{graur_supernovae_2011}, although any choice for $m_{\mathrm{Fe}}^{Ia}$ less than $\sim1M_{\odot}$ results in the same qualitative conclusions.
These results are shown in Figure \ref{fig:nia}.
Multiple studies based on SNIa surveys have converged to a volumetric rate of $N_{Ia}/M_{\star}\sim1\times10^{-3}~\textrm{SNIa}~M_{\odot}^{-1}$ \citep{graur_supernovae_2011,maoz_nearby_2011,maoz_star_2017,wiseman_rates_2021}.
\cite{weinberg_scale_2024} found a similar rate using a GCE model applied to the Milky Way with a fixed core-collapse Fe yield, assuming equilibrium is reached.
By contrast, the measurement from DLEIY is about 5 times higher than these estimates.
It is more consistent with the estimates from galaxy clusters, the LMC/SMC, and empirical scaling relationships with mass/metallicity \citep{maoz_supernova_2010,maoz_star_2017, brown_relative_2019, johnson_binaries_2023}.

We suggest that both of these results---a steep DTD and a high rate---could be a consequence of a metallicity dependence.
The close binary fraction of solar-type stars has been observed to increase significantly as metallicity decreases \citep{badenes_stellar_2018,moe_close_2019}, which, all else equal, is expected to both increase the number of SNeIa and increase the fraction that explode promptly \citep{greggio_rates_2005}.
Lower metallicity stars are also more likely to be in hierarchical triple systems, which can increase the eccentricity of the inner system and significantly decrease the merger time \citep{thompson_accelerating_2011}.
For the same initial stellar mass, a lower metallicity progenitor is also expected to produce a higher mass white dwarf.
Such progenitors should be easier to explode, which would increase the rate and decrease the merger time \citep{kistler_impact_2013}.




This picture is consistent with previous observational work that has found an inverse relationship between the specific rate and galaxy stellar mass \citep[e.g.,][]{li_nearby_2011,brown_relative_2019,wiseman_rates_2021}, although see \cite{graur_unified_2015,burgaz_ztf_2025}.
Specifically, \cite{brown_relative_2019} found a scaling of $N_{Ia}/M_{\star}\sim M_{\star}^{-0.5}$ using the ASAS-SN sample.
\cite{gandhi_exploring_2022} also found that metallicity-dependent rates that scale as $N_{Ia}/M_{\star}\sim Z^{-0.5}$ significantly improved the agreement of the FIRE-2 simulations with the ASAS-SN rates and with the observed mass-metallicity relation.
\cite{johnson_binaries_2023} explore the origins of this scaling, and they find that a scaling with metallicity of $N_{Ia}/M_{\star}\sim Z^{-0.5}$ is highly consistent with the close binary fraction measured by \cite{moe_close_2019} 
(although whether the observed scaling can be fully or only partially explained by binarity alone depends on the normalization of the mass-metallicity relation).
This corresponds to a slightly shallower scaling with mass compared to \cite{brown_relative_2019}, $N_{Ia}/M_{\star}\sim M_{\star}^{-0.3}$, assuming a mass-metallicity relation as in \cite{zahid_universal_2014}.
The rate from DLEIY is consistent with this shallower scaling.

There are, of course, complications with this picture.
First, fundamental processes in binary evolution like common envelope evolution are not fully understood in any regime \citep{ivanova_common_2013}.
Metallicity likely has multiple, mixed effects on binary evolution that affect the various possible progenitors of SNeIa differently, with some promoting and others suppressing prompt SNeIa \citep[e.g.,][]{kobayashi_low-metallicity_1998,meng_effect_2011,thiele_applying_2023}.
There is also mixed evidence from SN surveys for a mass dependence, which may be due to systematic uncertainties and assumptions about the stellar mass function \citep{greggio_correlation_2019}.
Although \cite{gandhi_exploring_2022} find that the mass dependence exists regardless of the choice of stellar mass function or star formation history, no metallicity dependence was found in the Zwicky Transient Facility data release 2 sample \citep{burgaz_ztf_2025}.
Also, while metallicity effects are qualitatively expected to increase the fraction of prompt SNeIa (i.e., produce a steeper DTD), it is not clear that they are sufficient to produce a DTD similar to $\sim t^{-2}$.



In general, there is less evidence for (or against) a metallicity-dependent shape of the DTD than for a metallicity-dependent SNIa rate. 
Some other forms of the DTD from different progenitors predict a more significant prompt component regardless of metallicity, such as the two-component model from \cite{mannucci_two_2006}.
The single-degenerate sub-Chandrasekhar scenario is expected to produce more prompt SNeIa as well \citep{greggio_rates_2005}.
\cite{ruiter_delay_2011} find that the DTD for a double-degenerate sub-Chandrasekhar mass scenario goes as $t^{-2}$ at longer delays ($t>1$ Gyr).
Indeed, some GCE studies have suggested that sub-Chandrasekhar mass white dwarfs may be the dominant progenitor for SNeIa in dwarf spheroidal galaxies with short star formation histories \citep{mcwilliam_evidence_2018,kirby_evidence_2019,de_los_reyes_manganese_2020}.
However, if this effect is indeed due to the duration of star formation (where galaxies with short SFHs can only incorporate material produced by prompt SNeIa, so prompt SNeIa are fractionally more important), we would not expect to see a steep DTD in Fornax.

The impact of star formation duration on observed stellar abundances is a broader issue that affects the interpretation of yields in all GCE models. 
The abundances of Sculptor are necessarily insensitive to very delayed SNeIa, because of the short duration of star formation ($\lesssim2$ Gyr).
Fornax has extended star formation, but is also not especially sensitive because most of its stars form at equilibrium.
Galaxies like Carina that have experienced a late burst of star formation could be informative, especially considering that the uncertainty in the observed SFHs is typically lower (and the time resolution finer) at shorter look-back times \citep{de_boer_episodic_2014}. 
A further way to address the contribution of different channels is to use abundances like Ni and Mn, which are sensitive to the mass of SNIa progenitor(s) \citep{seitenzahl_three-dimensional_2013}, and/or a more flexible form for the DTD.

It is also worth considering why these properties have not been identified previously, especially since the chemical evolution of Sculptor has been so extensively studied \citep[e.g.,][]{geisler_sculptor-ing_2005,kirby_multi-element_2009,de_boer_star_2012,hill_vltflames_2019, de_los_reyes_simultaneous_2022, tang_near-infrared_2023}.
Indeed, in previous work, a $t^{-1}$ distribution has largely held up, or at least, as been used without significant investigation.

Degeneracies with star formation (especially the star formation efficiency) may have confounded this issue.
Various combinations of the SFH and DTD can produce very similar abundance distributions (e.g., \citealt{palicio_cosmic_2024,dubay_galactic_2024}; see also \citealt{cote_impact_2017}).
Fixing the SFH and AZR significantly restricts the available parameter space for the DTD, and the only other knobs to turn are the abundances.
Simultaneously reproducing the observed abundances and CMD-based SFHs for both Sculptor and Fornax has also been challenging, suggesting that there is some mismatch of enrichment and star formation \citep[e.g.,][]{vincenzo_chemical_2014,hasselquist_apogee_2021,de_los_reyes_simultaneous_2022}.
Also, a high contribution from SNeIa relative to CCSNe in Fornax has been identified before as a possible explanation for its unusual chemical abundances relative to more massive satellites \citep{hasselquist_apogee_2021}.



\section{Conclusion}\label{sec:conclusion}

We have introduced a new galactic chemical evolution model called DLEIY, which uses observed SFHs and MDFs as inputs to constrain nucleosynthetic yields and the delay-time distribution of SNeIa.
We validate this method using simulated mock galaxies, and in particular show the sensitivity of DLEIY to the SNeIa delay-time distribution.
This method is applied to the chemical abundances and SFHs of the dwarf spheroidal galaxies Sculptor and Fornax using a joint statistical model of both galaxies, allowing information on environmentally-independent parameters to be shared.
We summarize the main results, with parameter values from the shared-yield+DTD model:
\begin{itemize}
    \item The joint model achieves simultaneous constraints on the population-averaged yields of Mg and Fe from CCSNe, Fe from SNeIa, and galactic outflows, without fixing the yield scale.
    \item The population-averaged CCSN Fe yield $y_{\mathrm{Fe}}^{CC}=1.17^{+0.37}_{-0.28}\times10^{-3}$ is in good agreement with theoretical models.
    \item The population-averaged CCSN Mg yield is a factor of a few higher than theoretical models, but in good agreement with previous literature that has found an underproduction of [Mg/Fe] in SN models \citep[e.g.,][]{philcox_optimal_2018}.
    \item Assuming an average Fe mass per SNIa of $m_{\mathrm{Fe}}^{Ia}=0.7M_{\odot}$, then the inferred specific rate of SNeIa, $N_{Ia}/M_{\star}$, is about 5 times higher than the typical rate for field galaxies. This aligns with previous work that has found an enhanced specific rate in low-mass galaxies \citep[e.g.,][]{brown_relative_2019}, which may be attributed to a metallicity dependence.
    \item A steep delay-time distribution ($\sim t^{-2}$) is preferred for both galaxies, in contrast to the fiducial distribution $\sim t^{-1}$ from most SNeIa surveys. This implies a much larger fraction of SNeIa must be prompt, rather than delayed. We suggest that this difference in the distribution may also be due to a metallicity dependence, where an increase in the close binary fraction and the final mass of white dwarfs enhances the rate of prompt SNeIa.
\end{itemize}

For some parameters, like the mass-loading factor, the uncertainties from DLEIY are quite large. 
These uncertainties may be appropriate in that they reflect the uncertainty in e.g., the SFH, but the level of imprecision does make it difficult to draw strong conclusions.
We expect that expanding this analysis to include more galaxies with diverse SFHs and masses would improve the precision, especially of the yields.
More precise star formation histories at long look-back times would also have a significant effect.
Deep, wide-field photometry expected from the Roman Space Telescope will be extremely valuable in this regard.
It would also resolve the mismatches between the area covered by SFHs and abundances.
DLEIY will also be well suited to the data from upcoming spectroscopic surveys like the Subaru Prime Focus Spectrograph Galactic Archaeology program \citep{takada_extragalactic_2014} and 4DWARFS \citep{skuladottir_4most_2023}.

\section{Acknowledgments}\label{sec:ack}
M.E.H, J.S.S, and T.S.L acknowledge the financial support of the Data Sciences Institute at the University of Toronto.
J.S.S was also supported by NSERC Discovery Grant RGPIN-2023-04849 and a University of Toronto Connaught New Researcher Award.
A.P.J. acknowledges support from the National Science Foundation under grants AST-2307599 and AST-2510795. A.P.J. acknowledges support from the Alfred P. Sloan Foundation.
This research benefited from the Dwarf Galaxies, Star Clusters, and Streams Workshop hosted by the Kavli Institute for Cosmological Physics.

We thank Erik Tollerud for devising the name DLEIY, and Steffani Grondin for valuable discussion regarding binary evolution.

\appendix
\section{Error Deconvolution of the MDF}\label{sec:app_decon}

To deconvolve uncertainty from the MDF, we use a hierarchical Bayesian model, in which our knowledge about the MDF informs the measured metallicities, and vice versa.

Consider a galaxy with some discrete, unobserved metallicity distribution $\boldsymbol{\rho}$, where each of $k$ bins has some probability $\rho_{k}$.
The metallicity of a random star in this galaxy, $Z$, can then be described by a categorical distribution.
A categorical distribution gives the probability of some discrete number of outcomes.
For example, a die has 6 possible outcomes, so a single roll of the dice is categorically distributed with $k=6$, and the probability of each outcome is $\boldsymbol{p}=\{\frac{1}{6},\frac{1}{6},\frac{1}{6},\frac{1}{6},\frac{1}{6},\frac{1}{6}\}$.
In our case, the probability of different outcomes (metallicities) $\boldsymbol{p}$ is simply $\boldsymbol{\rho}$, so the metallicity is distributed as $Z|\boldsymbol{\rho}\sim \text{Cat}(\boldsymbol{\rho})$, or equivalently, we can write $P(Z=k|\boldsymbol{\rho})= \rho_{k}$.

When we observe $N$ stars, we are making $N$ draws from this categorical distribution, which gives us a set of observed counts per bin $\boldsymbol{n}=\{n_{1},n_{2},...,n_{k}\}$.
Equivalently, we can think of drawing the entire set of observed counts per bin $\boldsymbol{n}$ from a distribution of possible realizations, each with some probability, i.e., from a multinomial distribution:

\begin{equation}\label{eq:n_true}
    \boldsymbol{n}|\boldsymbol{\rho} \sim \textrm{Mult}(N,\boldsymbol{\rho}) \iff 
    P(\boldsymbol{n}|\boldsymbol\rho) =
   \overbracket{\frac{N!}{n_1!n_2!\dots n_{k}!}}^{\mathclap{\substack{\text{Number of ways to put} \\ N~\text{observations in}~k~\text{bins}}}}
    \prod_{i=1}^{k}\rho_{i}^{n_{i}} 
    =\frac{\Gamma(\sum_{i=1}^{k}n_i+1)}{\prod_{i=1}^{k}\Gamma(n_i+1)}\prod_{i=1}^{k}\rho_{i}^{n_{i}}
\end{equation}

The multinomial distribution gives the probability of a certain set of counts per bin; if we throw our 6-sided die 10 times, the multinomial distribution gives the probability that we observe $\boldsymbol{n}=\{1,3,2,2,2,0\}$.

However, we are interested in the unobserved distribution $\boldsymbol{\rho}$ given the observed number of counts per bin, $P(\boldsymbol{\rho}|\boldsymbol{n})$, but what we have is $P(\boldsymbol{n}|\boldsymbol\rho)$, so we make use of Bayes' theorem.
For the prior $P(\boldsymbol{\rho})$, we choose a Dirichlet distribution, which is the multivariate generalization of the beta distribution:

\begin{equation}
   \boldsymbol{\rho}\sim\text{Dir}(\boldsymbol{\alpha})\iff  P(\boldsymbol{\rho})=
   \frac{\prod_{i=1}^{k}\Gamma(\alpha_{i})}{\Gamma(\sum_{i=1}^{k}\alpha_{i})}
   \prod_{i=1}^{k}\rho_{i}^{\alpha_{i}-1}
\end{equation}

The shape parameter $\boldsymbol{\alpha}$ that describes the Dirichlet distribution can be thought of as a pseudocount in each bin.
The Dirichlet is the conjugate prior of the multinomial, so the posterior $P(\boldsymbol{\rho}|\boldsymbol{n})$ is also Dirichlet distributed: $\boldsymbol{\rho}|\boldsymbol{n}\sim\text{Dir}(\boldsymbol{n}+\boldsymbol{\alpha})$.

In reality, we do not observe the true metallicities $Z$, but a noisy realization $Z_{obs}\sim\mathcal{N}(Z,\sigma^{2})$, so we do not know the true number of counts in each bin.
To handle this, we explicitly model the true, unobserved metallicity and iteratively update our guess for the unobserved metallicity of each object and the unobserved metallicity distribution function with Gibbs sampling.
Gibbs sampling is a Markov Chain Monte Carlo method that is used when sampling from the joint distribution is difficult, but sampling from the conditional distributions is possible \citep{geman_stochastic_1984}.
In Gibbs sampling, we sample from the conditional distribution of each variable in turn, updating the conditional distribution with the current value of the other variables.

We still model our guess for the true metallicity using a categorical distribution, but in this case, the probability of each bin is not $\boldsymbol{\rho}$.
Instead it is the overlap between our current guess for $\boldsymbol{\rho}^{\prime}$ and the likelihood of observing $Z_{obs}$ in a given bin, $P(Z_{obs}|Z,\sigma)$:
\begin{equation}\label{eq:cat_draw}
    Z|Z_{obs},\sigma,\boldsymbol{\rho}^{\prime}\sim \textrm{Cat}(\boldsymbol{p}),\textrm{~where~} \boldsymbol{p}=\frac{P(Z_{obs}|Z, \sigma)\odot\boldsymbol{\rho}^{\prime}}{P(Z_{obs}|Z, \sigma)\cdot\boldsymbol{\rho}^{\prime}}.
\end{equation}
Simply put, the probability that a star belongs to a given bin depends on how many stars should actually be in the bin ($\boldsymbol{\rho}^{\prime}$), and how close the measured metallicity is to that bin ($P(Z_{obs}|Z, \sigma)$).

Once our guess for the true metallicity of each star is updated, we draw a new population $\{Z^{\prime}\}$ with corresponding counts $\boldsymbol{n}^{\prime}$, which we use to update our guess for the unobserved metallicity distribution: $\boldsymbol{\rho}^{\prime}\sim \text{Dir}(\boldsymbol{n}^{\prime}+\boldsymbol{\alpha})$. 
This process is repeated, sampling in turn each true metallicity $\{Z\}$ and the underlying distribution $\boldsymbol{\rho}$. 
In Gibbs sampling, the samples for a single variable approximate its marginal distribution, which we use to draw realizations of the true, unobserved MDF.

We use a uniform shape parameter, $\alpha_{k}=1~\forall~k$.
Gibbs sampling is highly autocorrelated, so we thin the chain conservatively and keep only every 50th step. After thinning, 3500 steps are taken, with the first 1000 steps discarded as burn-in.
Neighboring bins are highly correlated on the scale of uncertainties, so we choose bins of width $0.15~\mathrm{dex}$ (median measurement uncertainties are $\sigma_{\mathrm{[Fe/H]}}\approx0.1$).
The maximum and minimum values of the bins and the bin width do affect the resulting distribution, but for a sensible choice, this method essentially always decreases bias in the AZR.

\section{Mock parameters and uncertainties}\label{sec:app_unc}
\setcounter{table}{0}
\renewcommand{\thetable}{A\arabic{table}}
\begin{deluxetable}{lcccccccccc}[h!]
\tablecaption{Fiducial parameters for the Sculptor-like and Fornax-like mocks used for validation.\label{tab:mocks}}
\tablehead{
\colhead{Model} &
\colhead{$y_{\mathrm{Mg}}^{\mathrm{CC}}$} &
\colhead{$y_{\mathrm{Fe}}^{\mathrm{CC}}$} &
\colhead{$y_{\mathrm{Fe}}^{\mathrm{Ia}}$} &
\colhead{$I_{0}$} &
\colhead{$\tau_{\mathrm{in}}$} &
\colhead{$\tau_{\star}$} &
\colhead{$M_{g,0}$} &
\colhead{$\eta$} &
\colhead{$\alpha$} &
\colhead{$t_{\min}$}
}
\startdata
Scl-like &
$3.5\times10^{-3}$ &
$1.2\times10^{-3}$ &
$4.5\times10^{-3}$ &
$3~M_{\odot}\,\mathrm{yr}^{-1}$ &
$1~\mathrm{Gyr}$ &
$12~\mathrm{Gyr}$ &
$10^{9}~M_{\odot}$ &
$50$ &
$-2.0$ &
$0.15$ \\
Fnx-like &
$3.5\times10^{-3}$ &
$1.2\times10^{-3}$ &
$4.5\times10^{-3}$ &
$1~M_{\odot}\,\mathrm{yr}^{-1}$ &
$10~\mathrm{Gyr}$ &
$8~\mathrm{Gyr}$ &
$0~M_{\odot}$ &
$40$ &
$-2.0$ &
$0.15$ \\
\enddata
\end{deluxetable}

Observed uncertainties in the SFH and abundances are heteroskedastic.
To approximate the SFH uncertainties in the mock tests, we add Laplace noise with a declining shape parameter as a function of time, $\beta(t)=\dot{M_{\star}}(t)e^{-t/10^5}$.
The Laplace distribution, also called the double-exponential distribution, is similar to the Gaussian distribution, but has much heavier tails. Qualitatively, this distribution better reproduces the observed uncertainties than a Gaussian distribution.

For abundance uncertainties, we assign uncertainties as a function of metallicity that approximate the observed uncertainties.
For the Fornax-like mock, the uncertainties are:
\begin{equation*}
    \begin{aligned}
        e_{[Fe/H]}\sim
        \begin{cases}
             U(0.1, 0.3)& \textrm{[Fe/H]} < -0.85~\\
             U(0.1, 0.15) & \textrm{[Fe/H]} \geq -0.85
        \end{cases}
    \end{aligned}
    \qquad\qquad
    \begin{aligned}
        e_{[Mg/Fe]}\sim
        \begin{cases}
             U(0.1, 0.5)& \textrm{[Fe/H]} < -0.85~\\
             U(0.1, 0.25) & \textrm{[Fe/H]} \geq -0.85
        \end{cases}
    \end{aligned}
\end{equation*}

For the Sculptor-like mock, we approximate the uncertainty as uniformly distributed between two exponential bounds as a function of metallicity, with a maximum/minimum uncertainty.
For [Mg/Fe], the lower bound of the uniform distribution is
$f_{low}(x)=\min{(0.1,10^{-6}e^{\frac{x - 2.7}{-0.4}}+0.08)}$, and the upper bound is $f_{up}(x)=\min{(0.5,10^{-6}e^{\frac{x - 2.7}{-0.4}}+0.2)}$.
For [Fe/H], the lower bound of the uniform distribution is $f_{low}(x)=\min{(0.1, 0.001e^{x - 1.3} +0.1)}$, and the upper bound is $f_{up}(x)=\min{(0.3,0.001e^{x - 0.9} +0.13)}$.

\newpage
\section{Corner plot for all parameters}\label{sec:app_corner}

Here we present the full corner plot for all parameters for the four models presented in this work.
The diagonal lines that appear in some parameter spaces for the shared-yield+DTD model arise because the DTD parameters ($\alpha$ and $t_{\mathrm{min}}$) are the same for both galaxies by definition.

\setcounter{figure}{0}
\renewcommand{\thefigure}{A\arabic{figure}}
\begin{figure}[h!]
\begin{center}
\includegraphics[width=\textwidth]{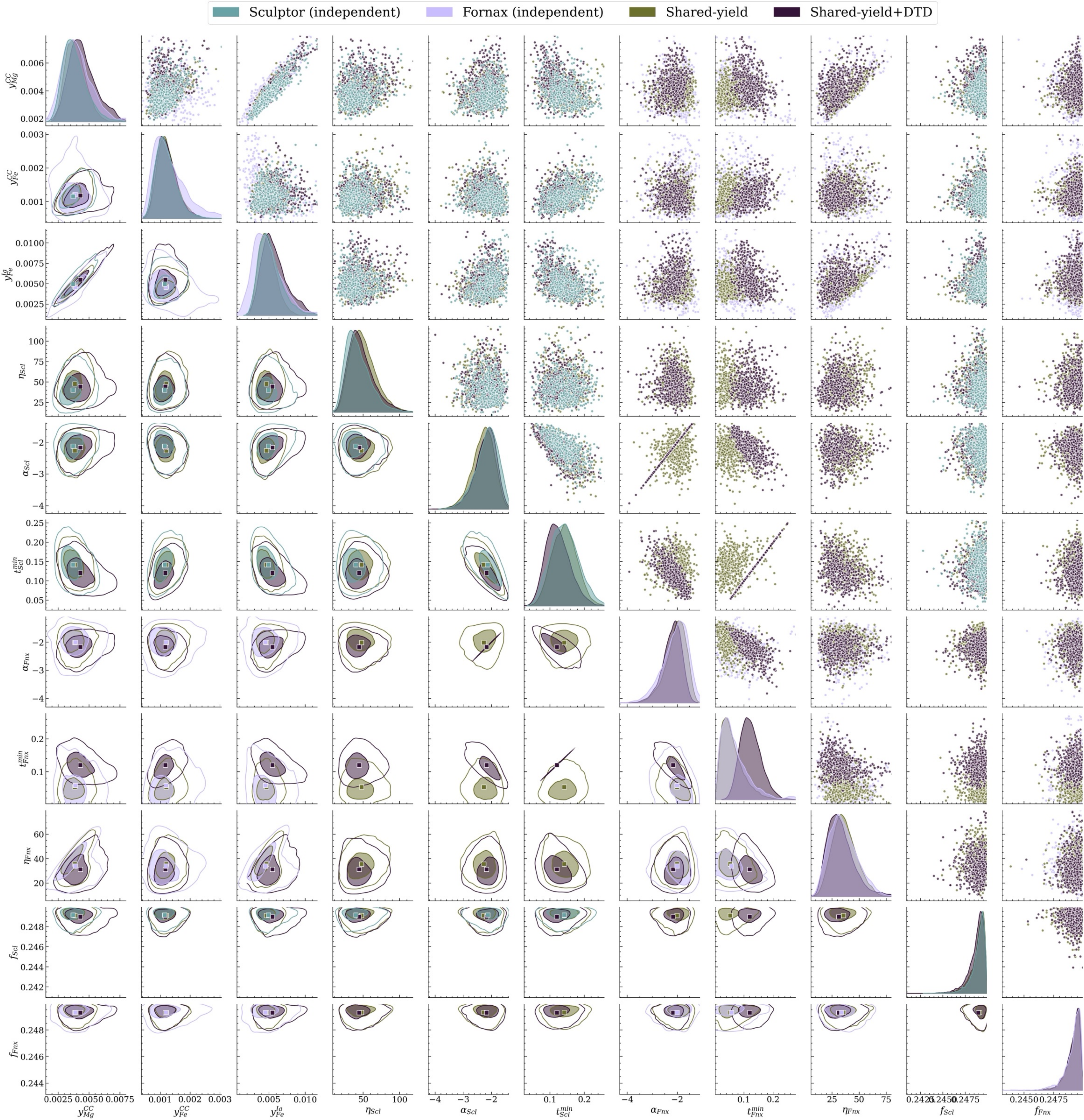}
\caption{Pair plot showing the correlations for all parameters. The contours on the lower diagonal are 40/86\%, approximately 1 and 2$\sigma$ in 2D. The point estimates are the median. On the upper diagonal, we show 500 posterior samples. \label{fig:corner_all}}
\end{center}
\end{figure}

\bibliography{dleiy_bib}{}
\bibliographystyle{aasjournal}

\end{CJK*}
\end{document}